# Optical transitions in hybrid perovskite solar cells: Ellipsometry, density functional theory and quantum efficiency analyses for $CH_3NH_3PbI_3$

Masaki Shirayama[1], Hideyuki Kadowaki[1], Tetsuhiko Miyadera[2], Takeshi Sugita[2], Masato Tamakoshi[1], Masato Kato[1], Takemasa Fujiseki[1], Daisuke Murata[1], Shota Hara[1], Takurou N. Murakami[2], Shohei Fujimoto[1], Masayuki Chikamatsu[2], and Hiroyuki Fujiwara[1*]

[1]Department of Electrical, Electronic and Computer Engineering, Gifu University, 1-1 Yanagido, Gifu 501-1193, Japan,
[2]Research Center for Photovoltaics, National Institute of Advanced Industrial Science and Technology (AIST), Central 5, 1-1-1 Higashi, Tsukuba, Ibaraki 305-8568, Japan.

## Abstract

Light-induced photocarrier generation is an essential process in all solar cells, including organic-inorganic hybrid ($CH_3NH_3PbI_3$) solar cells, which exhibit a high short-circuit current density ($J_{sc}$) of approximately 20 mA/cm$^2$. Although the high $J_{sc}$ observed in the hybrid solar cells relies on strong electron-photon interaction, the optical transitions in the perovskite material remain unclear. Here, we report artifact-free $CH_3NH_3PbI_3$ optical constants extracted from ultra-smooth perovskite layers without air exposure and assign all the optical transitions in the visible/ultraviolet region unambiguously based on density functional theory (DFT) analysis that assumes a simple pseudo-cubic crystal structure. From the self-consistent spectroscopic ellipsometry analysis of the ultra-smooth $CH_3NH_3PbI_3$ layers, we find that the absorption coefficients of $CH_3NH_3PbI_3$ ($\alpha$ = 3.8 × 10$^4$ cm$^{-1}$ at 2.0 eV) are comparable to those of $CuInGaSe_2$ and CdTe, and high $\alpha$ values reported in earlier studies are overestimated seriously by extensive surface roughness of $CH_3NH_3PbI_3$ layers. The polarization-dependent DFT calculations show that $CH_3NH_3^+$ interacts strongly with the $PbI_3^-$ cage, modifying the $CH_3NH_3PbI_3$ dielectric function in the visible region rather significantly. In particular, the transition matrix element of $CH_3NH_3PbI_3$ varies depending on the position of $CH_3NH_3^+$ within the Pb−I network. When the effect of $CH_3NH_3^+$ on the optical transition is eliminated in the DFT calculation, $CH_3NH_3PbI_3$ dielectric function deduced from DFT shows excellent agreement with the experimental result. As a result, distinct optical transitions observed at $E_0$ ($E_g$) = 1.61 eV, $E_1$ = 2.53 eV, and $E_2$ = 3.24 eV in $CH_3NH_3PbI_3$ are attributed to the direct semiconductor-type transitions at the R, M, and X points in the pseudo-cubic Brillouin zone, respectively. We further perform the quantum efficiency (QE) analysis for a standard hybrid-perovskite solar cell incorporating a mesoporous $TiO_2$ layer and demonstrate that the QE spectrum can be reproduced almost perfectly when the revised $CH_3NH_3PbI_3$ optical constants are employed. Depth-resolved QE simulations confirm that $J_{sc}$ is limited by the material's longer wavelength response and indicate the importance of optical confinement and long carrier diffusion lengths in hybrid perovskite solar cells.

*fujiwara@gifu-u.ac.jp

# I. INTRODUCTION

Since the first demonstration of methylammonium lead iodide ($CH_3NH_3PbI_3$) perovskite solar cells in 2009 [1] and the subsequent achievement of 10.9% conversion efficiency in 2012 [2], research into organic-inorganic hybrid solar cells has expanded dramatically [3−57], leading to a highest conversion efficiency now exceeding 20% [58]. One of the remarkable features of $CH_3NH_3PbI_3$ solar cells is the low carrier recombination at their interfaces; by simply introducing the perovskite light absorber layer between electron and hole transport layers, a high short-circuit current density ($J_{sc}$) of approximately 20 mA/cm$^2$ can be obtained with internal quantum efficiency (IQE) of 90–100% [4−6]. This is in sharp contrast to more conventional crystalline Si ($c$–Si) and $CuIn_{1−x}Ga_xSe_2$ (CIGS) solar cells, in which the use of a passivation or back-surface field layer is essential to reduce the strong interface recombination, particularly at the metal back contact [59]. In the organic-inorganic perovskite, the unique Pb–I bonding further suppresses the formation of midgap defect states [29−31]. The low carrier recombination within the $CH_3NH_3PbI_3$ bulk component is supported by observed long-range carrier diffusion lengths (100−1000 nm) [11,21−23], which can be attributed partly to small electron and hole effective masses ($m_e^* = 0.23m_0$ and $m_h^* = 0.29m_0$) [38] and the resulting high mobility (8−38 cm$^2$/Vs) [15,22,24,25] in $CH_3NH_3PbI_3$.

In ideal solar cells with low levels of bulk and interface recombination, $J_{sc}$ is ultimately controlled by the optical absorption in the solar cell absorber layer. Accordingly, determination and interpretation of $CH_3NH_3PbI_3$ optical properties are of critical importance for the further development of $CH_3NH_3PbI_3$ solar cells. It is known well that $CH_3NH_3PbI_3$ is a direct transition semiconductor with the fundamental transition at the R point (cubic symmetry) [28,30] or the Γ point (tetragonal and orthorhombic symmetries) [26,28,30] in the Brillouin zone. Nevertheless, for other optical transitions in the visible/ultraviolet region, the physical origins remain unclear. Specifically, in $CH_3NH_3PbI_3$, two intense absorption peaks have been observed at 2.5 eV and 3.4 eV. Earlier studies assumed that the optical absorption at 2.5 eV is caused by a transition from a second valence band to a conduction band [11,44], although the origin of the second band has not been explained. From the density functional theory (DFT), Even et al. proposed that the second optical transition occurs at the M point in the cubic phase [32,37]. However, the transition energy obtained from this DFT calculation (1.8 eV) is quite different from the experimental value of 2.5 eV. Later, Lin et al. attributed the 2.5-eV peak to a transition within a $PbI_2$ component remaining in $CH_3NH_3PbI_3$ [5]. More recently, this transition is proposed to occur from the valence band to the localized state formed within the conduction band [45,46]. Thus, the assignment of the 2.5-eV transition is highly controversial. For the optical transition at 3.4 eV, on the other hand, no assignment has been proposed.

There is also great uncertainty over the fundamental optical characteristics of the material [5–20]. In particular, the reported absorption coefficient $\alpha$ of $CH_3NH_3PbI_3$ differs significantly in a range from $2.5 \times 10^4$ cm$^{-1}$ [10] to $8.7 \times 10^4$ cm$^{-1}$ [6] at 2.0 eV. Light penetration though the absorber layer and the resulting carrier generation depend entirely on $\alpha$ of the absorber material and the above difference in $\alpha$ by a factor of three or more leads to a substantial variation in the device simulation of $CH_3NH_3PbI_3$–based solar cells. In addition, as the band gap ($E_g$) of $CH_3NH_3PbI_3$, slightly different values (1.50–1.61 eV) have been reported [14–20]. The above inconsistencies are likely to have been induced by non-ideal rough surface structures that are typically observed in solution-processed $CH_3NH_3PbI_3$ semiconductors [8,15]. Moreover, $CH_3NH_3PbI_3$ exhibits strong phase change in humid air [10, 47–51], and the previous optical studies may have been affected seriously by the $CH_3NH_3PbI_3$ degradation.

So far, extensive DFT calculations have been performed to reveal the optoelectronic properties of $CH_3NH_3PbI_3$ crystals [26–43]. However, the $CH_3NH_3PbI_3$ dielectric functions calculated from DFT [32–35] are different from the experimental dielectric functions reported so far [5–10]. In addition, the DFT-derived dielectric functions show strong anisotropic behavior when the tetragonal and orthorhombic structures are assumed [33,34]. Although the origin of the optical anisotropy has been attributed to the tetragonal distortion [34], no further details have been provided. More importantly, the perovskite phase structure that can approximate the room-temperature dielectric function accurately has not been specified. There has also been intensive research into the presence of an excitonic transition near the fundamental absorption edge [5,52,53,60–67]. Non-excitonic light absorption is consistent with the efficient free carrier generation within the absorber layer. Determination of the optical transition is therefore vital for a clear understanding of the operational principles of $CH_3NH_3PbI_3$ photovoltaic devices.

In this article, we present a general picture of the optical transitions in $CH_3NH_3PbI_3$ hybrid perovskite materials with the aid of DFT calculations. The origins of all optical transitions in the visible/ultraviolet region are explained on the basis of a simple pseudo-cubic crystal structure. We find that the optical anisotropy confirmed in the DFT calculation is caused by the strong interaction of $CH_3NH_3^+$ with the $PbI_3^-$ component. For accurate determination of the dielectric function, we characterize ultra-smooth $CH_3NH_3PbI_3$ layers, prepared by a laser evaporation technique, without exposing the samples to air and artifact-free optical data is obtained from multi-sample ellipsometry analysis [68]. As a result, the $CH_3NH_3PbI_3$ dielectric function determined from the experiment shows excellent agreement with the DFT result. We present evidence that the high $\alpha$ values reported in previous studies are overestimated seriously due to the effect of extensive roughness in $CH_3NH_3PbI_3$ samples. Optical device simulations performed using the revised $CH_3NH_3PbI_3$ optical constants further reproduce the $J_{sc}$ values observed experimentally in the high-efficiency perovskite solar cells.

## II. EXPERIMENT

The dielectric functions ($\varepsilon = \varepsilon_1 - i\varepsilon_2$) of $CH_3NH_3PbI_3$, $PbI_2$ and $CH_3NH_3I$ are evaluated by spectroscopic ellipsometry (SE). As known widely, SE is a surface-sensitive technique with roughness sensitivity on the atomic level [69], and preparation of samples with smooth surfaces is essential for reliable SE analysis. To prepare $CH_3NH_3PbI_3$ layers with ideal flat surfaces, we use a laser evaporation technique shown in Fig. 1(a). In conventional evaporation processes, control of the $CH_3NH_3I$ evaporation rate is rather difficult because of the high vapor pressure of this source, but the controllability is greatly improved by using the laser evaporation technique. In the laser evaporation process, $PbI_2$ (>99%, Aldrich) and $CH_3NH_3I$ (synthesized) source materials in crucibles are heated by a near-infrared laser with a wavelength of $\lambda = 808$ nm (Pascal, PA-LH-30LD). The evaporation rates of these materials are controlled by adjusting the power of the near-infrared laser. For $PbI_2$, the laser power intensity is adjusted to 3.2 W. In the case of $CH_3NH_3I$, the laser power of 17 W is modulated by a 10 Hz square wave for precise evaporation rate control. The resulting $CH_3NH_3PbI_3$ deposition rate is 0.6 nm/min. The laser evaporation process is conducted without substrate heating at a pressure of $5\times10^{-3}$ Pa. $CH_3NH_3PbI_3$ and $CH_3NH_3I$ layers are prepared on $c$−Si substrates that are coated with thin ZnO layers (50 nm) to improve film adhesion, whereas the $PbI_2$ layers are formed directly on the $c$−Si.

A scanning electron microscope (SEM) image of the $CH_3NH_3PbI_3$ layer confirms the formation of an ultra-smooth surface [Fig. 1(b)]. To avoid surface roughening and structural non-uniformity in the growth direction, the thin $CH_3NH_3PbI_3$ layer (45 nm) shown in Fig. 1(b) is characterized by SE. Despite the low thickness, this layer exhibits sharp x-ray diffraction (XRD) peaks that originate from the $CH_3NH_3PbI_3$ crystalline phase [3,4,28], and the $PbI_2$ diffraction peaks [4,70] are negligible in this film [Fig. 1(c)]. The XRD pattern of the $CH_3NH_3PbI_3$ in Fig. 1(c) is consistent with the formation of the cubic phase [28] and similar XRD spectra have been observed in the evaporated layers [3], although the XRD analysis of the single crystals confirms that the tetragonal phase is the most stable phase at room temperature [28]. Figure 1(d) summarizes atomic force microscopy (AFM) images of $CH_3NH_3PbI_3$, $PbI_2$ and $CH_3NH_3I$ layers used for the ellipsometry analyses. To perform reliable optical analysis, we employ two $CH_3NH_3PbI_3$ layers with thicknesses of 45 nm and 85 nm in Fig. 1(d) in the self-consistent SE analysis (see Sec. III). The root-mean-square roughness ($d_{rms}$) of $CH_3NH_3PbI_3$ (45 nm), estimated by AFM, is only 4.6 nm. In the case of the $CH_3NH_3I$ layer, however, the surface structure is quite non-uniform.

To perform optical simulation of $CH_3NH_3PbI_3$ solar cells, a $TiO_2$ electron transport layer, a *spiro*−OMeTAD [2,2',7,7'-tetrakis-(N,N-di-p-methoxyphenylamine) 9,9'-spirobifluorene] hole transport layer, a Ag layer and a $MoO_x$ layer are also prepared. The compact $TiO_2$ layer is deposited on a *c*−Si substrate with a native oxide layer by a spray pyrolysis method. Di-iso-propoxy titanium bis(acetylacetonate), as the precursor solution for compact $TiO_2$, is prepared by mixing 40 mmol of titanium tetraisopropoxide (>95.0%, Wako) and 80 mmol of acetylacetone (>99.0%, Wako) with a solvent of 2-propanol (>99.7%, Wako) for a 75 wt.% solution. The above precursor solution is further diluted with ethanol (>99.5% super dehydrated, Wako) for a 5% precursor solution and is sprayed on the *c*−Si substrate, which is heated to 450 °C, by the nebulizer. The $TiO_2$ layer thickness on the *c*−Si substrate is 10 nm.

For the *spiro*−OMeTAD layer, a spin-coating solution is prepared by dissolving 100 mg of *spiro*−OMeTAD (SHT-263 Livilux®, Merck) in 973 μl chlorobenzene with 9.6 μl of 4-tert-butylpyridine (Sigma-Aldrich) and 41 μl of lithium bis (trifluoromethylsulphonyl) imide (LiTFSI) acetonitrile solution (17 mg LiTFSI in 100 μl acetonitrile). The concentrations of *spiro*−OMeTAD, 4-tert-butylpyridine, and LiTFSI are 0.08 M, 0.06 M, and 0.02 M, respectively. The 40 μl spin-coating solution is dropped on a ZnO-coated *c*−Si substrate, and the substrate is then rotated at 2000 r.p.m. for 60 s. After coating, the substrate is heated at 100 °C to evaporate the solvent. The thickness of the *spiro*−OMeTAD layer on the substrate is 350 nm.

The Ag and $MoO_x$ layers are fabricated on glass substrates by conventional dc and rf sputtering at room temperature, respectively. In the case of $MoO_x$, we employ a pressure of 1.0 Pa, rf power of 40 W and a gas mixture of Ar and $O_2$ ($O_2$/Ar gas flow ratio is 0.05).

**III. SE ANLYSIS**

The SE spectra are measured using rotating-compensator instruments [69]. To avoid exposure of $CH_3NH_3PbI_3$ samples to humid air, the samples are sealed in a plastic bag inside a $N_2$-filled glove box attached to the laser evaporation system. The plastic bag with the samples is then transferred to a glove bag, in which the ellipsometry instrument is placed, and the plastic bag is opened after the glove bag is filled with $N_2$. From this procedure, the SE spectra of the pristine $CH_3NH_3PbI_3$ are obtained without exposing the samples to humid air. The SE measurements in $N_2$ are performed in a 0.7–5.2 eV energy range at an angle of incidence of 75° (J. A. Woollam M–2000XI). For ellipsometry characterization in air, we use another ellipsometer (J. A. Woollam, M–2000DI) that allows spectral measurements for up to 6.5 eV using various angles of incidence.

The SE analysis of the ultra-smooth $CH_3NH_3PbI_3$ layers is performed by using a global error minimization scheme [68]. In this approach, the dielectric function is determined self-consistently by using more than two samples with different layer thicknesses. For the $CH_3NH_3PbI_3$ analysis, two samples [45 and 85 nm in Fig. 1(d)] are used and we obtain two sets of ellipsometry spectra from these samples [i.e., $(\psi, \Delta)_{45nm}$ and $(\psi, \Delta)_{85nm}$]. In the self-consistent analysis, the dielectric function of the bulk layer $\varepsilon_{bulk}(E)$ is extracted directly from $(\psi, \Delta)_{45nm}$ using a mathematical inversion without assuming dielectric function models. In the second step, the deduced $\varepsilon_{bulk}(E)$ is applied to fitting analysis of $(\psi, \Delta)_{85nm}$. From this analysis procedure, $\varepsilon_{bulk}(E)$, the surface roughness layer thickness ($d_s$), and the bulk layer thickness ($d_b$) of the deposited layers can be determined, based on the assumption that the bulk layer optical properties are independent of the layer thickness [68]. For the SE analysis, we assume an optical model consisting of ambient/surface roughness layer/$CH_3NH_3PbI_3$ bulk layer/interface layer (3 nm)/ZnO (50 nm)/$SiO_2$ (2 nm)/$c$–Si substrate. The interface and $SiO_2$ layers correspond to the surface roughness of the ZnO layer and the native oxide of the $c$–Si substrate, respectively. The optical properties of the surface roughness layer are calculated as a 50:50 vol.% mixture of the bulk layer and voids by applying the Bruggeman effective-medium approximation (EMA) [69,71], while a 50:50 vol.% mixture of $CH_3NH_3PbI_3$ and ZnO is assumed for the interface layer.

Figure 2 shows experimental spectra of $(\psi, \Delta)_{85nm}$ and the solid lines represent the fitting result calculated using $\varepsilon_{bulk}(E)$ extracted from $(\psi, \Delta)_{45nm}$. It can be seen that the calculated spectra show excellent fitting to the experimental spectra. However, the SE fitting degrades slightly at $E \geq 3.0$ eV, probably due to the rough surface structure of the 85-nm-thick $CH_3NH_3PbI_3$ layer [see Fig. 1(d)]. In the self-consistent analysis, therefore, the fitting errors in the energy region of $E \leq 3.0$ eV are used, and we obtain $d_s = 2.9 \pm 0.1$ nm and $d_b = 40.6 \pm 0.1$ nm for $(\psi, \Delta)_{45nm}$ and $d_s = 9.4 \pm 0.1$ nm and $d_b = 76.1 \pm 0.1$ nm for $(\psi, \Delta)_{85nm}$. In the subsequent analysis, we extract the final $CH_3NH_3PbI_3$ dielectric function from ($\psi$, $\Delta$)$_{45nm}$ by adjusting $d_b$ slightly since the $\varepsilon_2$ values obtained from the above analysis show very small negative values ($\varepsilon_2 \sim -0.05$) at $E < E_g$. Thus, the $d_b$ value is increased slightly so that the $\varepsilon_2$ values at $E < E_g$ become completely zero. This $d_b$ adjustment is minor and corresponds to 6% of the total $d_b$. The $d_s$ of $2.9 \pm 0.1$ nm obtained from the above analysis shows reasonable agreement with the value of $d_{rms}$ = 4.6 nm observed in AFM, confirming the validity of the overall ellipsometry analysis [69].

We also determine the dielectric functions of air-exposed and thermal-annealed $CH_3NH_3PbI_3$ layers. For the $CH_3NH_3PbI_3$ layers exposed to humid air, we assume a constant $d_s$ of 2.9 nm estimated in the above analysis, whereas $d_s$ is treated as a free parameter in the SE analysis of the thermal-annealed samples. The $d_b$ values of the treated layers are determined from the SE fitting performed in a transparent region using the Tauc-Lorentz model [72]. By applying the $d_s$ and $d_b$ values, the dielectric functions of the air-exposed and thermal-annealed layers are extracted.

For $PbI_2$, a self-consistent SE analysis is also performed using two samples with thicknesses of 10 nm and 20 nm. The dielectric function of $PbI_2$ is obtained from the sample in Fig. 1(d) using the mathematical inversion. The $d_s$ of this $PbI_2$ layer ($1.4 \pm 0.1$ nm) also shows good agreement with the value of $d_{rms}$ = 0.9 nm observed in AFM.

In the ellipsometry analysis of the $CH_3NH_3I$ layer shown in Fig. 1(d), the optical model without the surface roughness layer (i.e., ambient/$CH_3NH_3I$ bulk layer/interface layer/ZnO/SiO$_2$/$c$−Si) is used, and the fitting analysis for the ($\psi$, $\Delta$) spectra is conducted in a transparent region without significant light absorption ($E \leq 3$ eV) using the Tauc-Lorentz model. From a $d_b$ value of 32.8 nm obtained in this analysis, the dielectric function is extracted using the mathematical inversion. Although the $CH_3NH_3I$ layer shows a rather rough surface structure, the effect of the surface roughness is negligible in the $CH_3NH_3I$ analysis. In particular, when $\varepsilon$ values are small, the roughness effect is averaged out over a whole layer as the difference between the dielectric functions of surface roughness and bulk layers becomes small [69].

For the TiO$_2$ and *spiro*−OMeTAD layers, ellipsometry analyses similar to that performed for $CH_3NH_3I$ are conducted. For the Ag layer, the dielectric function is deduced from the ($\psi$, $\Delta$) spectra while assuming a single optical layer without surface roughness. The dielectric function of a MoO$_x$ layer is extracted by a mathematical inversion assuming a simple model of ambient/surface roughness(void 50 vol.%)/bulk layer/substrate. In the mathematical inversion, the layer thicknesses determined by the Tauc-Lorentz analysis in a low energy region ($E \leq 4$ eV) are employed.

## IV. DFT CALCULATION

The DFT calculations of $CH_3NH_3PbI_3$ are implemented using a plane-wave ultrasoft pseudopotential method (Advance/PHASE software). For the exchange-correlation functional, the generalized gradient approximation (GGA) within the Perdew-Burke-Ernzerhof (PBE) scheme [73] has been applied. We perform structural optimization of a $CH_3NH_3PbI_3$ cubic structure using a $6 \times 6 \times 6$ k mesh and a plane-wave cutoff energy of 500 eV until the atomic configuration converged to within 5.0 meV/Å. The dielectric functions are calculated based on a method developed by Kageshima et al. [74]. For this calculation, we use a more dense $10 \times 10 \times 10$ k mesh to suppress distortion of the calculated spectra.

Previously, the importance of spin-orbit coupling in the DFT calculation has been pointed out [36,37]. Nevertheless, if this interaction is incorporated into the DFT calculation within PBE or the local density approximation (LDA), the $E_g$ reduces significantly down to 0.5 eV [36, 38–41], and the agreement with the experimental result degrades seriously. When the effect of spin-orbit coupling is considered using the *GW* approximation [39,40] or a hybrid functional [41], the experimental $E_g$ of 1.6 eV can be reproduced. However, the band structures obtained from these sophisticated calculations are essentially similar to that deduced from the simple PBE calculation [30]. Thus, in this study, the PBE calculation is performed without incorporating the effect of spin-orbit coupling. The same approach is employed in recent DFT studies [30,43].

## V. RESULTS

### A. Dielectric function of $CH_3NH_3PbI_3$

Figure 3(a) shows the dielectric functions of $CH_3NH_3PbI_3$, $PbI_2$ and $CH_3NH_3I$ obtained from the SE analyses in Sec. III. As described earlier, SE measurements of the $CH_3NH_3PbI_3$ layers are carried out in a $N_2$ ambient without exposing the samples to air. However, the dotted lines for $CH_3NH_3PbI_3$ at $E \geq 4.75$ eV are measured in air (relative humidity 40%) and are obtained within 20 s after air exposure. The SE measurements performed in air at various angles of incidence show that $CH_3NH_3PbI_3$ has isotropic optical properties. In Fig. 3(a), $\varepsilon_1$ for $CH_3NH_3PbI_3$ at $E = 0.75$ eV is 5.1 and is an intermediate level between the values of 7.1 for $PbI_2$ and 1.8 for $CH_3NH_3I$. The dielectric function of $PbI_2$ in Fig. 3(a) is similar to that reported earlier [75], although our result shows much higher $\varepsilon_2$ values at high energies ($E \sim 4$ eV).

Figure 3(b) presents the $\alpha$ spectra obtained from the dielectric functions of Fig. 3(a). $CH_3NH_3PbI_3$ shows a sharp onset of light absorption at $E_g \sim 1.6$ eV. The Urbach energy ($E_U$) of $CH_3NH_3PbI_3$, which is estimated by assuming $\alpha \propto \exp(E/E_U)$ [76], is 14 meV. This low $E_U$ confirms the suppressed tail state formation in $CH_3NH_3PbI_3$, as reported previously [13]. $PbI_2$ is a direct transition semiconductor [77,78], although the $\alpha$ values near $E_g \sim 2.0$ eV are very low [75,79]. A sharp optical transition at 2.5 eV in $PbI_2$ has been reported to be excitonic [78,80] and the transition energy of this peak is close to that of a $CH_3NH_3PbI_3$ peak observed in a similar energy region. However, the $PbI_2$ absorption peak at 2.5 eV is quite sharp, while the $CH_3NH_3PbI_3$ peak is very broad, suggesting that the origins of these transitions are different. On the other hand, light absorption in $CH_3NH_3I$ is quite weak and we observe four distinct transitions at 3.0 eV, 4.6 eV, 5.3 eV, and 6.2 eV in the $\alpha$ spectrum.

From the $\alpha$ spectrum of $CH_3NH_3PbI_3$ shown in Fig. 3(b), a conventional $E_g$ analysis is performed using a $(\alpha E)^2$–$E$ plot. Figure 3(c) presents the results of the $E_g$ analyses obtained using two different $(\alpha E)^2$ regions. As shown in Fig. 3(c), this analysis gives different values that depend on the energy regions analyzed and is unreliable. In our case, the $E_g$ values in a range of 1.58–1.63 eV are obtained. In order to estimate the optical transition energies accurately, critical point (CP) analysis is performed. In the CP analysis, the second-derivative $\varepsilon_1$ and $\varepsilon_2$ spectra are analyzed using the following theoretical formulas:

$$d^2\varepsilon/dE^2 = j(j-1)A\exp(i\phi)(E - E_p + i\Gamma)^{j-2} \quad (j \neq 0), \qquad (1)$$
$$d^2\varepsilon/dE^2 = A\exp(i\phi)(E - E_p + i\Gamma)^{-2} \qquad (j = 0), \qquad (2)$$

where $A$, $\phi$, $E_p$, and $\Gamma$ are the amplitude, phase, position and width of the peak, respectively [81]. Depending on the band structure, the CP is classified into one dimension ($j = -1/2$), two dimensions ($j = 0$) or three dimensions ($j = 1/2$). If an optical transition is excitonic, $j = -1$ is used in Eq. (1).

In the second-derivative spectra, however, the spectral noise is often enhanced drastically, which prevents accurate determination of the CP energies. For the CP analysis of $CH_3NH_3PbI_3$, therefore, the $CH_3NH_3PbI_3$ dielectric function is modeled using the Tauc-Lorentz model (see Fig. 10) and the second-derivative spectra are then calculated from the modeled dielectric function, as implemented previously [82,83]. The open circles in Fig. 3(d) show the $d^2\varepsilon_1/dE^2$ spectrum obtained from this procedure, while the solid line indicates the spectrum calculated from Eqs. (1) and (2). In this analysis, different combinations of the $j$ values are used to minimize the fitting error and the best fit is obtained when we use $j = 0$ for the $E_0$ transition and $j = -1$ for the $E_1$ and $E_2$ transitions. However, $j = -1/2$ also provides similar fitting quality for both the $E_1$ and $E_2$ transitions with almost identical CP energies, and further discussion is difficult. As a result, from this CP analysis, the $E_0$ ($E_g$) of $CH_3NH_3PbI_3$ is found to be 1.61 ± 0.01 eV, while the peak transition energies in Fig. 3(a) are determined to be $E_1$ = 2.53 ± 0.01 eV and $E_2$ = 3.24 ± 0.01 eV.

Figure 4(a) compares the $\varepsilon_2$ spectrum determined from our analysis to those reported earlier in other studies [5–10]. It can be seen that, although the spectral features are rather similar, the absolute values and the transition energies are quite different. In particular, the $E_2$ peak position is shifted to 3.4 eV in Refs. [6,8–10] and the amplitude of the $E_2$ transition is smaller in Refs. [6–9]. Figure 4(b) summarizes the $\alpha$ spectra of $CH_3NH_3PbI_3$ reported previously [5–13]. As confirmed from this figure, the absolute values of $\alpha$ differ significantly. In Fig. 4(b), the $\alpha$ spectra are denoted by dotted lines when the $\alpha$ spectrum shows relatively large values in the energy region below $E_g$.

We find that the large variations in the dielectric functions and the $\alpha$ spectra in Figs. 4(a) and (b) can be explained by two effects of (i) hydrate phase formation near the surface region and (ii) large surface roughness in $CH_3NH_3PbI_3$ samples. Our SE analysis shows that the degradation of $CH_3NH_3PbI_3$ in humid air alters the optical spectrum in a short-wavelength region significantly. Figure 4(c) presents the change of the $CH_3NH_3PbI_3$ $\varepsilon_2$ spectrum in air at 40% relative humidity, and the $\varepsilon_2$ spectrum denoted as "in $N_2$" corresponds to the optical data in Fig. 3(a). When the $\varepsilon_2$ spectrum is characterized in a $N_2$ ambient, the amplitudes of the $E_1$ (2.53 eV) and $E_2$ (3.24 eV) transition peaks are high. However, as the exposure time to air increases, the peak values reduce gradually and the $E_2$ peak shifts toward higher energies. The peak observed at 3.4 eV in humid air is consistent with the formation of a hydrated crystal phase: i.e., $(CH_3NH_3)_4PbI_6 \cdot 2H_2O$ [60]. In the previous studies [5–10], the particular control of the measurement environment was not made. Accordingly, the $E_2$ peak position at 3.4 eV observed in the earlier studies suggests the presence of the hydrated phase on the surface. Since the penetration depth of the light probe is small at high energy due to high $\alpha$ (19 nm at 3.4 eV), the optical response is more affected by a near-surface structure in this region. It should be emphasized that the phase change occurs more rapidly at higher relative humidity and, in our case, the $CH_3NH_3PbI_3$ surface becomes whitish within 20 s in air at 75% relative humidity. Thus, the optical characterization of $CH_3NH_3PbI_3$ needs to be performed without exposing samples to air. The detail of the $CH_3NH_3PbI_3$ degradation in humid air will be reported elsewhere.

Some of previous results are also influenced strongly by extensive roughness of $CH_3NH_3PbI_3$ samples. It is now well established that effective medium theories are valid when $D < 0.1\lambda$, where $D$ and $\lambda$ are the dimension of microstructures and the wavelength of light probe, respectively [71,84]. To apply EMA, therefore, dimensions of surface roughness need to be much smaller than $\lambda$ of SE measurements. In solution-processed $CH_3NH_3PbI_3$ layers, the lateral size of crystal grains is around 300 nm [8,15] and the $D$ is comparable to $\lambda$. In this condition ($D \sim \lambda$), the surface roughness cannot be represented by a single EMA layer and a complex EMA-multilayer model is necessary to express the optical response in the rough surface region properly [82,85,86]. When the SE analysis is performed using an oversimplified optical model, the extracted dielectric function generally shows strong artifacts, such as non-zero $\varepsilon_2$ and $\alpha$ values in the energy region even below $E_g$.

To reveal the effect of underestimated roughness contribution, we calculate pseudo-dielectric functions ($\langle\varepsilon\rangle = \langle\varepsilon_1\rangle - i\langle\varepsilon_2\rangle$) from our $CH_3NH_3PbI_3$ dielectric function assuming a hypothetical surface roughness layer. Figure 4(d) shows the variation of the $\langle\varepsilon_2\rangle$ spectrum and the pseudo-$\alpha$ spectrum ($\langle\alpha\rangle$) with the roughness layer thickness in a range of 0–25 nm, calculated from an optical model of ambient/surface roughness/$CH_3NH_3PbI_3$ substrate. In this simulation, the experimental dielectric function in Fig. 3(a) is used and the optical response of the roughness layer is modeled by EMA using a 50:50 vol.% mixture of the bulk component ($CH_3NH_3PbI_3$) and voids. In the calculation of the $\langle\alpha\rangle$ spectrum, we first obtain $\langle k\rangle$ from the corresponding $\langle\varepsilon\rangle$ and the $\langle k\rangle$ spectrum is then converted to $\langle\alpha\rangle$ by $\langle\alpha\rangle = 4\pi\langle k\rangle/\lambda$. When the roughness thickness is zero, $\langle\varepsilon_2\rangle$ and $\langle\alpha\rangle$ are equivalent to $\varepsilon_2$ and $\alpha$, respectively.

The calculation result of Fig. 4(d) indicates the notable increase in $\langle\varepsilon_2\rangle$ and $\langle\alpha\rangle$ at low energies, while the amplitude of the $\langle\varepsilon_2\rangle$ peaks at high energies reduces sharply with constant peak energies. It can be seen that the $\varepsilon_2$ spectra of Refs. [6, 7] in Fig. 4(a) are reproduced quite well if $\langle\varepsilon_2\rangle$ is calculated from our $\varepsilon_2$ spectrum assuming an additional void contribution in the surface region. As mentioned earlier, when $D \sim \lambda$, the void component within the roughness region cannot be expressed by a simple EMA roughness layer and the remaining void fraction is incorporated into the bulk component. Accordingly, if we introduce the extra void fraction to our data, the calculated $\langle\varepsilon_2\rangle$ shows a close match with the $\varepsilon_2$ spectrum obtained assuming an oversimplified model. On the other hand, when the ellipsometry analysis is performed using dielectric function models, the $\varepsilon_2$ component below $E_g$ is forced to be zero. Therefore, the serious disagreement between the $\varepsilon_2$ spectra, observed particularly among spin-coated $CH_3NH_3PbI_3$ layers [6–9], can be attributed to the effect of large surface roughness. In Ref. [5], an evaporated film was characterized in air and the reported $\varepsilon_2$ is rather similar to our result, whereas the $\varepsilon_2$ spectrum obtained from the single crystal [10] shows a slightly different shape.

The large variation in the reported $\alpha$ spectra in Fig. 4(b) can also be explained by the roughness effect in $CH_3NH_3PbI_3$ samples. Specifically, the simulated $\langle\alpha\rangle$ spectra in Fig. 4(d) reproduce the high $\alpha$ values reported in Refs. [6, 7] quite well. In the case of transmission measurements, if transmission loss induced by surface light scattering is interpreted as light absorption, this analysis generates persistent $\alpha$ values even below $E_g$. Thus, the $\alpha$ values denoted by the dotted lines are likely overestimated seriously and these spectra indeed show quite high $\alpha$ values, compared with the other results, except for the spectra of Refs. [10, 11]. In Ref. [13], on the other hand, only the effective $\alpha$ values of a solution-processed $CH_3NH_3PbI_3$ layer were estimated by neglecting the effects of optical interference and light scattering. If the above $\alpha$ spectra are excluded, the other results show similar values of $4 \times 10^4$ cm$^{-1}$ at 2.0 eV. When the $\alpha$ values are overestimated, the $E_g$ value is in turn underestimated in a conventional $E_g$ analysis [see Fig. 3(c)], as confirmed from Figs. 4(b) and (d). Thus, the variation in the reported $E_g$ values (1.50–1.61 eV in Refs. [14–20]) can be attributed partly to the roughness effect, although the uncertainty in the $E_g$ analysis is also large.

In Fig. 4(b), our $\alpha$ spectrum shows the lowest values near the $E_g$. We rule out the possibility that the low $\alpha$ values are caused by a void-rich structure, as the $\varepsilon_2$ (or $\alpha$) values at high energies are higher than those of the other studies and the film structure is highly uniform [Fig. 1(b)]. To justify that the $\alpha$ spectrum obtained in this study shows the inherent optical properties of the $CH_3NH_3PbI_3$ perovskite, we have further performed thermal annealing of the room-temperature deposited $CH_3NH_3PbI_3$ layers shown in Fig. 1. The annealing experiments are performed at 80 and 100 °C for 45 min, as implemented previously for evaporated $CH_3NH_3PbI_3$ layers [3]. In our experiment, however, the annealing was carried out under vacuum with and without the $CH_3NH_3I$ vapor, rather than under $N_2$ atmosphere [3], in an attempt to realize better control of near-surface structures.

Figure 5 shows (a) XRD, (b) SEM, (c) dielectric function, and (d) $\alpha$ spectra of the $CH_3NH_3PbI_3$ layers annealed at 80 °C and 100 °C for 45 min in vacuum without the $CH_3NH_3I$ flux. In this figure, the results of the as-deposited $CH_3NH_3PbI_3$ layer shown in Figs. 1 and 3 are also indicated. It can be seen from Fig. 5(a) that the XRD spectrum shows little change after the annealing at 80 °C, but the crystal orientation is more random after the annealing, as confirmed from the appearance of the small diffraction peaks at $2\theta = 24.5°$ and $31.8°$. In contrast, the 100-°C annealing leads to the dominant formation of the $PbI_2$ phase due to the desorption of $CH_3NH_3I$-related species. The presence of $PbI_2$ cannot be eliminated even when the $CH_3NH_3I$ vapor is supplied during the 100-°C annealing at least in our conditions.

The SEM images in Fig. 5(b) show the significant change in the surface structure upon thermal annealing. Specifically, after the 80-°C annealing, the ultra-smooth surface of the as-deposited layer changes to the rough surface due to the formation of larger $CH_3NH_3PbI_3$ grains with sizes of 100–200 nm. The surface morphology of this sample is rather similar to that obtained from conventional solution-processed $CH_3NH_3PbI_3$ [8,15], although the grain size of our sample is still smaller. The 100-°C annealing leads to the smaller grains with non-uniform coverage of the underlying ZnO layer. Thus, the variation of the thermal annealing temperature in this regime has a large impact on the resulting structure.

From the SE analysis of the annealed layers, we obtained $(d_s, d_b)$ = (7.9 nm, 38.5 nm) for 80 °C and $(d_s, d_b)$ = (4.6 nm, 32.6 nm) for 100 °C. The $d_{rms}$ values obtained from the AFM measurements are 7.1 nm (80 °C) and 4.9 nm (100 °C) and show excellent agreement with the $d_s$ values determined in SE. The $\varepsilon_2$ and $\alpha$ spectra in Figs. 5(c) and (d) indicate clearly that, in the visible region ($E < 2.5$ eV), the optical properties of the 80-°C annealed $CH_3NH_3PbI_3$ are almost identical to those of the as-deposited $CH_3NH_3PbI_3$. From this result, we conclude that the $\alpha$ spectrum of Fig. 3(b) shows an intrinsic property of $CH_3NH_3PbI_3$ hybrid perovskite crystals and does not change significantly with crystal grain size.

In the 80-°C annealed layer, however, the amplitude of the dielectric function reduces at $E > 2.5$ eV. We attribute this to the effect of the extensive roughness in the annealed layer, as the condition of $D < 0.1\lambda$ is no longer satisfied for this sample. The slight increase in the $\alpha$ below $E_g$, observed for the 80-°C annealed layer, also indicates the underestimated roughness contribution, as confirmed from the simulation result of Fig. 4(d). When the $CH_3NH_3PbI_3$ is annealed at 100 °C, the visible light absorption disappears almost completely by the formation of the $PbI_2$ phase. It can be seen that the spectral shape of the 100-°C annealed layer is quite similar to that of $PbI_2$ shown in Fig. 3(a).

The above results indicate that the ultra-smooth $CH_3NH_3PbI_3$ layers fabricated by laser evaporation at room temperature are ideal samples particularly for accurate determination of the optical constants of $CH_3NH_3PbI_3$. We re-emphasize that, in our case, the $CH_3NH_3PbI_3$ samples are characterized in a $N_2$ ambient and the consistent optical results have been obtained in the multi-sample analysis. The validity of our optical constants is further confirmed from excellent agreement between experimental and calculated external quantum efficiency (EQE) spectra for an actual solar cell (see Sec. V C).

## B. Optical transitions in CH$_3$NH$_3$PbI$_3$

To determine the origin of the light absorption in CH$_3$NH$_3$PbI$_3$, we implement DFT calculations of CH$_3$NH$_3$PbI$_3$ assuming a simple pseudo-cubic structure. This assumption is critical to reproduce the optical transitions in the hybrid perovskite as described below. From structural optimization of the CH$_3$NH$_3$PbI$_3$ crystal within the cubic basis, we obtain the lattice parameters $a = 6.306$ Å, $b = 6.291$ Å, and $c = 6.310$ Å with $\alpha = \gamma = 90.00°$ and $\beta = 90.35°$ for the unit cell [Fig. 6(a)]. These lattice parameters are comparable to the experimental value in the literature ($c = 6.22$–$6.32$ Å) [17,28]. In the calculated structure, the PbI$_6$ octahedron is tilted and the C–N bond is aligned to be almost parallel to the $a$ axis [27,33,40], although other CH$_3$NH$_3^+$ configurations are also energetically possible [26–43].

Figure 6(b) compares the experimental $\varepsilon_2$ spectrum to the $\varepsilon_2$ spectra that are calculated for the different polarization states with directions parallel to the $a$, $b$ and $c$ axes in Fig. 6(a). In Fig. 6(b), the $\varepsilon_2$ spectra in two different energy regions are shown. DFT analysis reveals the highly anisotropic optical properties of CH$_3$NH$_3$PbI$_3$, and the dielectric function for the $a$ axis polarization ($\varepsilon_a$) is quite different to those for the $b$ and $c$ axes ($\varepsilon_b$ and $\varepsilon_c$). Nevertheless, the overall shapes of $\varepsilon_b$ and $\varepsilon_c$ show remarkable agreement with the shape of the experimental spectrum. Reasonable agreement is also observed for the $\varepsilon_2$ spectra in the $E_0$ transition regime. Therefore, the pseudo-cubic structure in Fig. 6(a) reproduces the optical transitions in CH$_3$NH$_3$PbI$_3$ surprisingly well. However, the DFT calculation shows that $\alpha = 6.2 \times 10^4$ cm$^{-1}$ at 2.0 eV ($\varepsilon_b$), which is notably larger than the experimental value of $\alpha = 3.8 \times 10^4$ cm$^{-1}$.

Figure 6(c) presents the band structure and the density of states (DOS) calculated from the pseudo-cubic structure of Fig. 6(a). In this figure, the partial DOS of the Pb 6$s$, Pb 6$p$ and I 5$p$ states is also shown. For the band structure, the corresponding Brillouin zone is shown in Fig. 6(d). V$_j$ and C$_j$ in Fig. 6(c) denote the $j$th valence and conduction bands from the valence band maximum (VBM) and the conduction band minimum (CBM), respectively. The band structure and DOS in Fig. 6(c) are essentially similar to those reported earlier [26,28–31,33,38–41] and V$_1$ consists of the Pb 6$s$ and I 5$p$ states, whereas C$_1$ is dominated by Pb 6$p$ [26–30]. Because of the antibonding nature of Pb–I [87], the charge densities are localized at the VBM and CBM [27], as confirmed from the inset of Fig. 6(c). In the band structure of Fig. 6(c), the energy positions of M$_{1-3}$ and those of X$_{1-3}$ differ slightly because the assumed cubic crystal is distorted and the resulting reciprocal lattices ($a^* = 2\pi/a$, $b^* = 2\pi/b$, $c^* = 2\pi/c$) are different.

To understand the optical transitions and the anisotropic optical behavior, we calculate the dielectric response induced by each interband transition. In particular, we perform detailed analyses for transitions of $V_j \to C_j$ ($j \leq 4$) to cover the energy region of $E \leq 5$ eV. Figure 7 presents contribution of each interband transition to the $\varepsilon_2$ spectrum of $CH_3NH_3PbI_3$, and the $\varepsilon_2$ spectra denoted as "Total" correspond to the $\varepsilon_2$ spectra in Fig. 6(b). In Fig. 7, $V_1C_2$ indicates the $\varepsilon_2$ contribution induced by the optical transition from the first valence band ($V_1$) to the second conduction band ($C_2$), for example. In this figure, however, only the transitions with a peak amplitude of $\varepsilon_2 > 0.5$ are shown for clarity. From the band structure in Fig. 6(c) and the $\varepsilon_2$ contribution in Fig. 7, we select the optical transitions at high symmetry points that satisfy van Hove singularities [88] in k space: i.e., $\nabla_\mathbf{k} [E_c(\mathbf{k}) - E_v(\mathbf{k})] = 0$, where $E_c(\mathbf{k})$ and $E_v(\mathbf{k})$ show the energies of the conduction and valence bands, respectively. From this analysis, we find that the dielectric response can be categorized by the reciprocal lattice and $\varepsilon_b$ is derived mainly from the $b^*$ component (R, $M_1$, $M_2$, and $X_2$ points). In the $V_1C_1$ transition, a small spike at 2.5 eV, which is attributed to the transition at the $M_2$ point, can be seen for $\varepsilon_b$ and $\varepsilon_c$, whereas this peak is absent in the $V_1C_1$ transition of $\varepsilon_a$. Since the $M_2$ point consists of the $b^*$ and $c^*$ components [see Fig. 6(d)], the $\varepsilon_2$ peak at 2.5 eV is present only for $\varepsilon_b$ and $\varepsilon_c$. The optical transitions determined by such analyses are indicated by the arrows in Fig. 6(c) and the transition energies are $E_0 = 1.56$ eV, $E_1 = 2.46$ eV, $E_{2,a} = 3.20$ eV, $E_{2,b} = 3.29$ eV and $E_{2,c} = 3.33$ eV. These transitions represent those for $b^*$, and similar transitions also occur for $c^*$. The CP energies indicated in Fig. 6(b) correspond to the above transition energies. As a result, the $E_1$ and $E_2$ transitions are attributed to direct transitions at the M and X points, respectively. When assuming the cubic symmetry, the direct $E_0$ transition occurs at the R point, as reported previously [28,30].

As confirmed from Fig. 7, the optical transition in the visible region is dominated by the $V_1C_1$ transition for all the polarization states. However, the calculated dielectric functions show large optical anisotropy at $E = E_1$. The transition analysis in Fig. 7 shows clearly that the optical anisotropy is generated by the difference in the $V_1C_2$ transition, because the $V_1C_1$ contribution at $E = E_1$ is rather independent of the polarization state. As shown in Fig. 6(c), the $V_1C_2$ transition at $E = E_1$ occurs at the M point and the charge density contour of 2.0–2.2 eV, which corresponds to the $E_1$ transition, shows distinct charge localization on the N atom (see the dotted circle). This result implies that the strong optical anisotropy observed in the DFT-derived dielectric function is caused by the interaction of the N atom with the $PbI_3^-$ cage.

To reveal the optical effect of the N atom in the DFT calculation more precisely, the dielectric functions are calculated by replacing $CH_3NH_3^+$ with $NH_4^+$. As reported previously [40], the band structure of $NH_4PbI_3$ is quite similar to that of $CH_3NH_3PbI_3$.

Figure 8(a) shows the NH$_4$PbI$_3$ dielectric function deduced from DFT. In this calculation, we assume that (i) the PbI$_3^-$ atomic configuration in NH$_4$PbI$_3$ is exactly same with that in CH$_3$NH$_3$PbI$_3$ shown in Fig. 6(a) and (ii) NH$_4^+$ locates at the center position of the C−N bond in the CH$_3$NH$_3$PbI$_3$. When this hypothetical NH$_4$PbI$_3$ structure is assumed, all the dielectric functions (i.e., $\varepsilon_a$, $\varepsilon_b$ and $\varepsilon_c$) show a similar shape and the strong optical anisotropy disappears. The small remaining anisotropy is caused by the orientation of NH$_4^+$ and the $\varepsilon_2$ increases slightly when the N−H axis is parallel to the polarization direction.

In the following calculation, NH$_4^+$ in Fig. 8(a) is relaxed while fixing the atomic configuration of the PbI$_3^-$ cage and, in this case, NH$_4^+$ moves along the *a* axis by 0.7 Å due to a strong interaction with the PbI$_3^-$ [Fig. 8(b)]. In earlier molecular dynamics (MD) simulations of CH$_3$NH$_3$PbI$_3$ crystals, the strong coupling between the NH$_3$ group and the I atom has also been confirmed [89]. In particular, the MD calculation reveals that the I and H of the NH$_3$ group show a stabilized distance of 2.65 Å [89], which explains the behavior of NH$_4^+$ in Fig. 8(b) well. Moreover, the N atom position of NH$_4^+$ in Fig. 8(b) is almost identical to that of CH$_3$NH$_3^+$ shown in Fig. 6(a). Quite surprisingly, the dielectric functions obtained with the relaxed NH$_4^+$ show strong anisotropic behavior and the calculation result is almost the same with that of CH$_3$NH$_3$PbI$_3$ in Fig. 6(b). Thus, the optical anisotropy changes significantly with the NH$_4^+$ position within the Pb−I network and the effect of CH$_3$ group is negligible. In fact, when the N atom of CH$_3$NH$_3^+$ is shifted to the same N atom position of the NH$_4$PbI$_3$ in Fig. 8(a), we obtain the dielectric functions that are almost identical to those of Fig. 8(a). Accordingly, a small change in the N atom configuration induces a large difference in $\varepsilon_a$ and the strong optical anisotropy is eliminated even in the case of CH$_3$NH$_3$PbI$_3$.

The effect of the Pb−I−Pb bond angle on the dielectric function is also investigated. In Fig. 8(c), the CH$_3$NH$_3$PbI$_3$ dielectric function is calculated assuming the Pb−I−Pb bond angle of 180$^\circ$ in the optimized CH$_3$NH$_3$PbI$_3$ structure of Fig. 6(a). It can be seen that the optical anisotropy reduces in this structure, compared with that of the original structure shown in Fig. 6(a). Nevertheless, our systematic dielectric-function calculations with different configurations indicate that the anisotropic character of CH$_3$NH$_3$PbI$_3$ and NH$_4$PbI$_3$ is primarily governed by the distance between N and I atoms. In particular, the reduction of the optical anisotropy in Fig. 8(c) can be interpreted by a larger N−I atom distance, compared with the case of Fig. 6(a), in which the Pb−I−Pb bond angle along the *b* axis is 169$^\circ$ and the N−I atom distance is slightly closer. When the Pb−I−Pb bond angle is 180$^\circ$, on the other hand, the onset of $\varepsilon_2$ shifts down to 1.34 eV due to the reduction in $E_g$. Our DFT calculations reveal that the Pb−I−Pb bond angle modifies the DOS near the CBM and the $E_g$ increases at larger Pb–I bond bending.

The above results show clearly that the N atom interacts rather strongly with the $PbI_3^-$ and modifies the dielectric function in the visible region. In the $NH_4PbI_3$ and $CH_3NH_3PbI_3$ structures shown in Fig. 8, on the other hand, the DOS distributions do not vary significantly. In addition, for the $E_1$ transition, the partial DOS distributions are dominated by the Pb and I contributions, and those of the N, C and H atoms are negligible. These results indicate that the optical anisotropy in $CH_3NH_3PbI_3$ originates from the change in the transition matrix element and is induced through the indirect interaction of $CH_3NH_3^+$. As mentioned above, the optical anisotropy derived from the DFT calculations is one dimensional. Unfortunately, this one-dimensional anisotropy is rather difficult to explain from the confirmed dependence of the optical properties on the N−I atom distance, and further detail is not clear at this stage.

As described in Sec. V A, the optical properties of $CH_3NH_3PbI_3$ observed experimentally are isotropic. Theoretically, the dielectric function corresponding to $(\varepsilon_a + \varepsilon_b + \varepsilon_c)/3$ is expected for the polycrystalline phase. However, the observed $\varepsilon_2$ spectrum is different from the $\varepsilon_2$ deduced from $(\varepsilon_a + \varepsilon_b + \varepsilon_c)/3$ shown in Fig. 7(d) and indicates the small contribution of $\varepsilon_a$ or the weak involvement of $CH_3NH_3^+$ in the optical transition. This contradiction can be interpreted through the reorientation of $CH_3NH_3^+$ in the $PbI_3^-$ cage [90–93]. In particular, the nuclear magnetic resonance shows ultra-fast reorientation of the C–N bond in $CH_3NH_3PbI_3$, with a relaxation time of < 0.5 ps [90]. Therefore, the results shown in Fig. 6(b) present evidence that the interaction between $CH_3NH_3^+$ and $PbI_3^-$ is hindered strongly by the extremely rapid reorientation of $CH_3NH_3^+$ at room temperature. In DFT calculations that assume a temperature of 0 K, the $CH_3NH_3^+$ position is completely fixed, and the optical spectrum is modified greatly by $CH_3NH_3^+$. Thus, to reproduce the dielectric function observed experimentally at room temperature, the effect of $CH_3NH_3^+$ needs to be minimized intentionally.

When the MD simulation is performed for $CH_3NH_3PbI_3$, on the other hand, $CH_3NH_3^+$ reorients rapidly with the relaxation time of the order of 1 ps at room temperature and the quasi-random orientation of $CH_3NH_3^+$ within the $PbI_3^-$ cage can be reproduced [93]. If such a dynamical rotation is considered and the dielectric function is calculated as a weighted average of different $CH_3NH_3^+$ orientations, the dielectric function at room temperature can be predicted more accurately, although this approach is computationally quite demanding.

Figure 9(a) summarizes the $\varepsilon_2$ spectra of $CH_3NH_3PbI_3$ calculated in previous DFT studies [32–35], together with our experimental and DFT spectra. In this figure, $\varepsilon_b$ in Fig. 6(b) is shown as the DFT dielectric function in this study. The earlier DFT results have been obtained assuming the cubic [32], tetragonal [33,35] and orthorhombic [34] structures. As reported previously, the band structure is basically independent of the crystal structure [28,30,42]. The anisotropic optical properties have also been confirmed in the tetragonal [33] and orthorhombic [34] structures. For these studies, only $\varepsilon_a$ spectra (*a* axis component in tetragonal and orthorhombic unit cells) are shown in Fig. 9(a), because the $\varepsilon_c$ spectra (*c* axis component) are quite different from the experimental result.

In Fig. 9(a), although the overall shapes of the dielectric functions are similar, the energies and structures of the critical points are rather different. Specifically, in all the previous calculations, the $E_0$ transition is not reproduced well and the $\varepsilon_2$ value in this region shows only a gradual increase with $E$. We find that the absence of the clear $E_0$ transition originates from the change in the DOS. Figure 9(b) shows the DOS distributions of the pseudo-cubic and tetragonal structures. For the DOS calculation of the tetragonal crystal, a reported DFT-optimized structure [35] is assumed. The DOS of the pseudo-cubic structure in Fig. 9(b) has been shown in Fig. 6(c). It can be seen that the DOS of the pseudo-cubic phase shows a sharp cliff-like feature, which corresponds to the $E_0$ transition, whereas the DOS of the tetragonal phase indicates a smoothened distribution near the CBM. Thus, the poor agreement with the experimental dielectric function in the $E_0$ transition regime, confirmed in the reported tetragonal and orthorhombic structures, is likely caused by the change in the DOS distribution near the CBM. In addition, the interaction with $CH_3NH_3^+$ also reduces the amplitude of the $E_0$ transition, as confirmed from the $\varepsilon_a$ spectrum of Fig. 6(b).

For the $E_2$ transition, the CP energies of the previous studies are slightly lower. This small shift can be attributed to the interaction of $CH_3NH_3^+$. In fact, the dielectric function of the tetragonal structure [33] in Fig. 9(a) is reproduced well if the $\varepsilon_a$ and $\varepsilon_b$ components of the pseudo-cubic structure in Fig. 6(b) are mixed, as the *a*, *b* axes of the tetragonal structure are rotated by 45° with respect to the cubic unit cell. In the tetragonal and orthorhombic structures, various $CH_3NH_3^+$ configurations are possible and the interpretation of the optical transitions becomes difficult. On the other hand, the CP energies obtained from the pseudo-cubic structure show good agreement with the experimental result. Accordingly, it is essential to assume the cubic phase in the DFT calculation to reproduce the experimental dielectric function properly. The MD calculation also indicates rapid $CH_3NH_3^+$ reorientations in the tetragonal and cubic phases [93]. Thus, the optical effect of $CH_3NH_3^+$ needs to be minimized in both cases.

In this study, the DFT calculations are implemented by PBE without considering electron-hole interactions or exciton formation. Thus, the excellent overall agreement between the experimental and theoretical dielectric functions reflects the important nature of optical absorption in $CH_3NH_3PbI_3$; all the optical transitions (i.e., $E_{0-2}$) are essentially non-excitonic. In the solar cells, the optical absorption in the visible region ($E_0 \leq E \leq E_1$) is important and variation of the light absorption in this regime is characterized by a featureless non-excitonic interband transition in the Brillouin zone. Accordingly, light absorption in $CH_3NH_3PbI_3$ can be represented by conventional semiconductor-type optical transitions and free electrons and holes are generated directly by light absorption at $E \geq E_g$.

The non-excitonic nature of the fundamental $E_0$ transition in $CH_3NH_3PbI_3$ has already been reported previously [5,66,67]. Specifically, the very small exciton binding energies of ≤ 6 meV at room temperature have been confirmed based on the determination of the static dielectric constant [5], the theoretical analysis of the absorption spectra [66] and the magneto-optical characterization [67]. On the other hand, a distinct excitonic transition is observed experimentally in the orthorhombic phase formed at low temperatures (≤ 170 K) [52,60-62,64,66,67] with the exciton binding energy of 16−50 meV [61,62,64,66,67]. Although our calculation does not take the excitonic effect into account, the excitonic transition observed in the orthorhombic $CH_3NH_3PbI_3$ can be reproduced well when the DFT calculation is performed by solving Bethe-Salpeter equation [65].

**C. Optical simulation of perovskite solar cells**

By using the revised $CH_3NH_3PbI_3$ optical constants determined in this study, the optical simulation of the hybrid solar cells is performed. Figure 10 summarizes the dielectric functions of the solar-cell component layers ($CH_3NH_3PbI_3$, $TiO_2$, *spiro*–OMeTAD, and Ag), together with the $PbI_2$ and $CH_3NH_3I$ layers, extracted from the SE analyses described in Sec. III. Similar dielectric functions have already been reported for anatase $TiO_2$ [94] and *spiro*–OMeTAD [6]. To eliminate the spectral noise, the dielectric functions in Fig. 10 are modeled using the Tauc-Lorentz model [72]. In this model, the $\varepsilon_2$ spectrum is expressed as a product of the Tauc optical gap and the Lorentz model:

$$\varepsilon_2(E) = \frac{ACE_p(E-E_g)^2}{(E^2-E_p^2)^2+C^2E^2} \cdot \frac{1}{E} \quad (E > E_g) \;, \tag{3}$$

$$= 0 \quad (E \leq E_g) \;, \tag{4}$$

where $A$, $C$, and $E_p$ represent the amplitude parameter, broadening parameter, and peak transition energy, respectively. The corresponding $\varepsilon_1(E)$ can be obtained from Eqs. (3) and (4) using the Kramers-Kronig relations:

$$\varepsilon_1(E) = \varepsilon_1(\infty) + \frac{AC}{\pi \xi^4} \cdot \frac{a_{\ln}}{2\beta E_p} \ln\left(\frac{E_p^2 + E_g^2 + \beta E_g}{E_p^2 + E_g^2 - \beta E_g}\right)$$
$$- \frac{A}{\pi \xi^4} \cdot \frac{a_{\tan}}{E_p} \left[\pi - \tan^{-1}\left(\frac{2E_g + \beta}{C}\right) + \tan^{-1}\left(\frac{-2E_g + \beta}{C}\right)\right]$$
$$+ 2\frac{AE_p}{\pi \xi^4 \beta} E_g (E^2 - \gamma^2) \left[\pi + 2\tan^{-1}\left(2\frac{\gamma^2 - E_g^2}{\beta C}\right)\right] \qquad (5)$$
$$- \frac{AE_p C}{\pi \xi^4} \cdot \frac{E^2 + E_g^2}{E} \ln\left(\frac{|E - E_g|}{E + E_g}\right) + \frac{2AE_p C}{\pi \xi^4} E_g \ln\left[\frac{|E - E_g|(E + E_g)}{\sqrt{(E_p^2 - E_g^2)^2 + E_g^2 C^2}}\right]$$

where

$$a_{\ln} = (E_g^2 - E_p^2)E^2 + E_g^2 C^2 - E_p^2(E_p^2 + 3E_g^2), \qquad (6)$$

$$a_{\tan} = (E^2 - E_p^2)(E_p^2 + E_g^2) + E_g^2 C^2, \qquad (7)$$

$$\xi^4 = (E^2 - \gamma^2)^2 + \beta^2 C^2/4, \qquad (8)$$

$$\beta = \sqrt{4E_p^2 - C^2}, \qquad (9)$$

$$\gamma = \sqrt{E_p^2 - C^2/2}. \qquad (10)$$

In Eq. (5), $\varepsilon_1(\infty)$ shows a constant contribution to $\varepsilon_1(E)$ at high energies. In the actual $\varepsilon_1(E)$ calculation, $E$ values should be chosen so that $|E - E_g|$ in Eq. (5) does not become zero. As a result, the dielectric function of the Tauc-Lorentz model is described by five parameters [$A$, $C$, $E_g$, $E_p$, $\varepsilon_1(\infty)$].

The dielectric functions of the Ag and SnO$_2$:F layers are modeled by combining the Tauc-Lorentz model with the Drude model: i.e., $\varepsilon = -A_D/(E^2 - i\Gamma_D E)$ [69]. The free carrier absorption in transparent conductive oxides can be expressed completely from two Drude parameters defined by the optical carrier concentration ($N_{opt}$) and optical mobility ($\mu_{opt}$) [95]. Based on earlier studies [96,97], we assume $N_{opt} = 2 \times 10^{20}$ cm$^{-3}$ and $\mu_{opt} = 30$ cm$^2$/(Vs) in the SnO$_2$:F layer. Table I summarizes all the parameters extracted from the dielectric function modeling in Fig. 10. The $\alpha$ spectrum of CH$_3$NH$_3$PbI$_3$ shown in Fig. 3(b) can be expressed quite well from the modeling parameters of CH$_3$NH$_3$PbI$_3$ in Table I.

Using the above optical data, the EQE spectrum of a standard hybrid perovskite solar cell, consisting of a glass/SnO$_2$:F (600 nm)/TiO$_2$ (200 nm)/CH$_3$NH$_3$PbI$_3$ (400

nm)/*spiro*−OMeTAD (500 nm)/Ag structure, is calculated. For the glass substrate, the effect of the multiple light reflection/transmission is neglected. In the EQE simulation, the absorptance spectrum for $\lambda$ [i.e., $A(\lambda)$] in each layer is estimated first using a conventional optical admittance method [98−100] while assuming a flat structure. The EQE spectrum is then deduced from $A(\lambda)$ of the perovskite layer assuming 100% collection efficiency of generated carriers [100].

Figure 11 shows the calculated $A$ and EQE spectra of the hybrid perovskite solar cell. The reflectance ($R$) spectrum of the solar cell is also shown and, due to the assumed flat structure, rather strong interference effect appears in the whole $\lambda$ region. The EQE spectrum of the $CH_3NH_3PbI_3$ indicates high EQE values of approximately 85% in the visible region. In the $SnO_2$:F, the strong light absorption occurs at $E \geq 3.3$ eV ($\lambda \leq 375$ nm) due to the interband transition, whereas the light absorption at $\lambda > 375$ nm indicates the contribution by the free carrier absorption. The $TiO_2$ layer shows the interband transition at $\lambda < 385$ nm. Since the $E_g$ of *spiro*−OMeTAD (2.95 eV) is higher than that of $CH_3NH_3PbI_3$, the light absorption in this layer is negligible. The $J_{sc}$ of the solar cell shown in Fig. 11 is 19.9 mA/cm$^2$ under AM1.5G conditions. Accordingly, $J_{sc} \sim 20$ mA/cm$^2$ can be reproduced quite well from our optical constants with a $CH_3NH_3PbI_3$ layer thickness of 400 nm. The optical simulation shows a reflection loss of 3.6 mA/cm$^2$ and parasitic optical losses of 1.3 mA/cm$^2$ ($SnO_2$:F), 0.3 mA/cm$^2$ ($TiO_2$), and 0.1 mA/cm$^2$ (Ag) in the energy region above $E_g = 1.61$ eV for $CH_3NH_3PbI_3$. When a natural texture is present, the front light reflection is suppressed and the $J_{sc}$ of the $CH_3NH_3PbI_3$ solar cell increases. If the above reflection loss is eliminated completely by the texture, the $J_{sc}$ of 23.5 mA/cm$^2$ can be obtained.

From the result of Fig. 11, the IQE and internal absorptance (IA) spectra are calculated according to IQE = EQE/(1 − $R$) and IA = $A$/(1 − $R$). Figure 12(a) presents the calculated IQE and IA of the solar cell. It can be seen that the IQE maximum in the range of $\lambda$ = 390−720 nm is limited by free carrier absorption in the $SnO_2$:F layer. To visualize carrier generation within the solar cell, partial IQE and IA spectra are calculated further by dividing the $CH_3NH_3PbI_3$ and $TiO_2$ layers into 400 and 200 sublayers, respectively, as implemented previously [100]. In Fig. 12(b), the partial IQE and IA calculated for different depths from the interfaces and wavelengths are shown. These values are normalized relative to the maximum value in each layer. If the partial IQE and IA spectra obtained at different depths are integrated, those shown in Fig. 12(a) are obtained. The partial IQE of the $CH_3NH_3PbI_3$ layer in the short-wavelength regime is limited by light absorption in the upper $TiO_2$ layer. At $E \geq E_1$, the partial IQE exhibits rapid decay with increasing depth because of strong light absorption within $CH_3NH_3PbI_3$. In contrast, in the region of $E_0 \leq E < E_1$, the IQE sensitivity is low because of the smaller $\alpha$ values and photocarriers are generated uniformly throughout the entire $CH_3NH_3PbI_3$ layer with the appearance of the optical interference effect. In this region, the electrons and holes that are generated near the $CH_3NH_3PbI_3$/*spiro*−OMeTAD and the $TiO_2$/$CH_3NH_3PbI_3$ interfaces, respectively, need to travel through the whole $CH_3NH_3PbI_3$ layer. Therefore, the high $J_{sc}$ of 20 mA/cm$^2$ observed experimentally supports the long-range carrier diffusion length, which is at least comparable to the $CH_3NH_3PbI_3$ layer thickness. In Fig. 12(b), the integrated $J_{sc}$ values relative to the depth from the $TiO_2$/$CH_3NH_3PbI_3$ interface and $\lambda$ are shown. The variation of $J_{sc}$ with $\lambda$ reproduces a previously reported trend [4]. The contribution of $J_{sc}$ at $\lambda \geq 500$ nm accounts for 73% of the total $J_{sc}$, confirming that the longer wavelength response is critical to realize a high $J_{sc}$ [5].

Figure 12(c) presents the change in $J_{sc}$ with $CH_3NH_3PbI_3$ thickness. The solid lines show the integrated $J_{sc}$, similar to that shown in Fig. 12(b), which is estimated using $CH_3NH_3PbI_3$ layers with discrete thicknesses ranging from 100 to 1000 nm, and the end points are shown by the solid circles. The dotted line in this figure shows the $J_{sc}$ values obtained by varying the $CH_3NH_3PbI_3$ layer thickness in the solar cell structure. Previously, the $J_{sc}$ dependences on the $CH_3NH_3PbI_3$ layer thickness were also reported using the different $\alpha$ spectra of $CH_3NH_3PbI_3$ [5,6]. In Fig. 12(c), the difference in $J_{sc}$ shown between the dotted line and the solid line for the 1000 nm thickness indicates the influence of the back-side reflection and the improvement of $J_{sc}$ by the back-side reflection is 2.4 mA/cm$^2$ at a layer thickness of 400 nm. Thus, we find that the contribution of the rear reflection is relatively large in the hybrid solar cell. This effect can be seen more clearly in IQE simulation results obtained from three solar cell structures with $CH_3NH_3PbI_3$ thicknesses of 200, 400 and 600 nm (Fig. 13). The simulation result of 400 nm in Fig. 13 corresponds to the partial IQE in Fig. 12(b). When the $CH_3NH_3PbI_3$ thickness is 200 nm, the partial IQE values at $\lambda > 500$ nm are higher than those of the thicker $CH_3NH_3PbI_3$ layers, indicating that the light is absorbed more effectively in the thin layer by enhanced back-side reflection. Accordingly, the optimum thickness of 400 nm confirmed in Fig. 12(c) can be understood as a consequence of the optical confinement in the hybrid solar cells.

To justify the $\alpha$ values of $CH_3NH_3PbI_3$ obtained in this study and the above EQE simulations, we have further performed the EQE analysis for an experimental perovskite solar cell fabricated by a standard solution process. In conventional hybrid solar cells having metal back contacts, however, the effect of the back-side reflection is strong and the confirmation of the absolute $\alpha$ values is more difficult. Accordingly, we analyze the EQE spectrum of a $CH_3NH_3PbI_3$ solar cell developed originally for a 4-terminal tandem solar cell consisting of $CH_3NH_3PbI_3$ top and CIGS bottom cells [101]. In the $CH_3NH_3PbI_3$ top cell of this device, transparent conductive oxide (TCO) layers are employed as the front and rear electrodes with a Ni-Al metal grid electrode on the rear side. In this case, the optical absorption in the solar cell is essentially determined by a single optical pass within the $CH_3NH_3PbI_3$ layer due to the weak back-side reflection. Thus, this "semi-transparent solar cell" provides an ideal test structure to validate the overall optical response within the solar cell.

The structure of the reported $CH_3NH_3PbI_3$ top cell consists of $MgF_2$/glass/$SnO_2$:F/compact $TiO_2$ (30 nm)/mesoporous $TiO_2$–$CH_3NH_3PbI_3$ (150 nm)/$CH_3NH_3PbI_3$ (240 nm)/*spiro*–OMeTAD/$MoO_x$/ZnO:Al/(Ni-Al grid)/$MgF_2$ [101]. For this solar cells, the conversion efficiency of 12.1% with the $J_{sc}$ of 16.7 mA/cm$^2$, open-circuit voltage of 1.03 V and fill factor of 0.703 is reported [101]. Figure 14(a) shows the optical model constructed for this top cell. The layer thicknesses in the optical model were extracted from the description and SEM image in Ref. [101]. For the $SnO_2$:F, however, a $SnO_2$:F/$SiO_2$/$SnO_2$ structure on the glass substrate (TEC glass) [102] is treated as a single $SnO_2$:F layer with a thickness of 600 nm. We confirm the validity of this assumption from the optical simulation. More importantly, to simplify the optical modeling, the optical response within the mesoporous $TiO_2$–$CH_3NH_3PbI_3$ mixed-phase layer (150 nm) was expressed by the two separate flat layers of a $TiO_2$ layer (60 nm) and a $CH_3NH_3PbI_3$ layer (90 nm) assuming a $TiO_2$ volume fraction of 40% (porosity of 60%), reported for the mesoporous $TiO_2$ layers processed using a commercial paste (Dyesol) [103,104]. As a result, in our optical simulation, the complex structure of the compact $TiO_2$ (30 nm)/mesoporous $TiO_2$–$CH_3NH_3PbI_3$ mixed phase (150 nm)/uniform $CH_3NH_3PbI_3$ (240 nm) is described as a simple two layer structure of $TiO_2$ (90 nm)/$CH_3NH_3PbI_3$ (330 nm) by neglecting the light scattering effect.

Figure 14(b) shows the optical constants (refractive index $n$ and extinction coefficient $k$) of the TCO and $MgF_2$ anti-reflecting layers used for the optical model of Fig. 14(a). For the other layers in the optical model [i.e., $CH_3NH_3PbI_3$, $TiO_2$, *spiro*–OMeTAD in Fig. 14(a)], the optical data in Fig. 10 is employed. The ($n$, $k$) spectra of the $MgF_2$, $SnO_2$:F (TEC-15) and ZnO:Al (sputtered film) correspond to those reported in Ref. [100], Ref. [102] and Ref. [100], respectively. The increase in $k$ at longer wavelengths, observed for the $SnO_2$:F and ZnO:Al in Fig. 14(b), shows the free carrier absorption and the Drude parameters of these layers are $A_D$ = 1.956 eV and $\Gamma_D$ = 0.085 eV [$N_{opt}$ = 4.5 × 10$^{20}$ cm$^{-3}$, $\mu_{opt}$ = 43 cm$^2$/(Vs)] for the $SnO_2$:F [102] and $A_D$ = 0.850 eV and $\Gamma_D$ = 0.114 eV [$N_{opt}$ = 1.8 × 10$^{20}$ cm$^{-3}$, $\mu_{opt}$ = 34 cm$^2$/(Vs)] for the ZnO:Al [100]. The carrier concentrations of the SnO$_2$:F layers in TEC substrates (TEC-8 and TEC-15), characterized by Hall measurements, are 5 × 10$^{20}$ cm$^{-3}$ [105,106] and agree quite well with the $N_{opt}$ value of the SnO$_2$:F, although the Hall mobility of the SnO$_2$:F [20–30 cm$^2$/(Vs)] is smaller than $\mu_{opt}$ due to the effect of grain boundary scattering [95]. For the MoO$_x$, the light absorption below $\lambda$ = 450 nm shows the interband transition, whereas the absorption peak at $\lambda$ = 925 nm ($E$ = 1.34 eV) indicates the optical transition associated with the defect band [107,108] and the amplitude of this peak reduces with increasing O$_2$/Ar gas flow ratio. We confirm that the effect of this defect absorption is negligible in our optical simulation.

The optical simulation of the semi-transparent solar cell is performed by using the optical admittance method assuming the flat structure. As known well, for the transmittance ($T$), $R$ and $A$, there is a relation of $T_{ex} + R_{ex} + A_{ex} = 1$, where the subscript "ex" shows the experimental spectrum. In our calculation, the $R_{ex}$ obtained from the actual solar cell is adopted to simulate the $A$ and $T$ more accurately. This method has been applied for accurate EQE simulation of CIGS solar cells [100].

Figure 14(c) shows the $T_{ex}$ and $R_{ex}$ spectra of the semi-transparent cell reported in Ref. [101] (open circles), together with the $T$ spectra calculated from the optical model of Fig. 14(a) (solid lines). As shown in Fig. 14(c), we calculated the $T$ spectra assuming different CH$_3$NH$_3$PbI$_3$ dielectric functions reported earlier. Specifically, the dielectric functions that show the highest $\alpha$ (Ref. [6]) and the moderate $\alpha$ (Ref. [8]) in Fig. 4 are employed to find the effect of the $\alpha$ on the $T$ spectrum. In the optical simulation for Ref. [6], however, we eliminated the nominal light absorption below $E_g$ (1.6 eV) by modeling the dielectric function using the Tauc-Lorentz model and this modified optical data was used. It can be seen that the $T$ spectrum calculated from our optical constants in Fig. 3 [red line in Fig. 14(c)] agrees quite well with the $T_{ex}$, whereas the calculated $T$ values deviate from $T_{ex}$ when the optical constants of Ref. [6] (green line) and Ref. [8] (blue line) are employed.

Figure 14(d) shows the EQE spectrum reported in Ref. [101] (open circles), together with the calculated EQE spectra obtained simultaneously in the above simulation (solid lines). The simulated EQE spectra are obtained directly from the calculated $A$ spectra assuming 100% collection of carriers generated within the CH$_3$NH$_3$PbI$_3$ layer. As confirmed from Fig. 14(d), the EQE spectrum calculated from our optical constants shows excellent agreement with the experimental EQE spectrum. This result indicates clearly that the $T$ and $A$ ($EQE$) spectra of a mesoporous TiO$_2$−CH$_3$NH$_3$PbI$_3$ solar cell fabricated by a standard solution process can be reproduced quite well from the simple optical simulation using the CH$_3$NH$_3$PbI$_3$ optical constants in Fig. 3. In contrast, when the $CH_3NH_3PbI_3$ optical constants with higher $\alpha$ values are employed in the calculation, the EQE (or light absorption within the $CH_3NH_3PbI_3$) is overestimated seriously particularly in the low $\alpha$ region ($E < E_1$ or $\lambda > 490$ nm) and the resulting $T$ decreases significantly, compared with the $T_{ex}$. Accordingly, the optical simulations in Figs. 14(c) and (d) strongly support the validity of the $CH_3NH_3PbI_3$ optical constants reported in this study.

In Fig. 14(d), the $J_{sc}$ value estimated from the integration of the calculated EQE spectrum at $\lambda > 350$ nm is indicated and the $J_{sc}$ obtained from our optical constants (16.5 mA/cm$^2$) agrees quite well with the experimental $J_{sc}$ of 16.7 mA/cm$^2$, although the almost perfect agreement originates in part from the cancellation of the slight disagreements at the shorter and longer wavelength regions. The optical absorption losses obtained from this EQE analysis are 2.3 mA/cm$^2$ ($SnO_2$:F), 0.04 mA/cm$^2$ ($TiO_2$), $3.5 \times 10^{-5}$ mA/cm$^2$ (*spiro*–OMeTAD), 0.03 mA/cm$^2$ ($MoO_x$) and 0.2 mA/cm$^2$ (ZnO:Al) with the reflection and transmission losses of 0.7 and 5.0 mA/cm$^2$, respectively, in the energy region above $E_g = 1.61$ eV ($CH_3NH_3PbI_3$). For this "semi-transparent solar cell", therefore, the transmission loss is rather significant.

It should be emphasized that the $J_{sc}$ is affected strongly by the EQE in the longer wavelength region as the number of photons increases in this region. Thus, although the EQE spectrum calculated using the optical data of Ref. [8] is rather close to the experimental EQE, the $J_{sc}$ is overestimated largely by 2 mA/cm$^2$. In addition, the $\alpha$ value reported in Ref. [6] is quite high and the calculated $J_{sc}$ shows a high value of 21.1 mA/cm$^2$. The result of Fig. 14(d) indicates clearly that the simulated EQE and the resulting $J_{sc}$ vary significantly with the $\alpha$ spectrum used for $CH_3NH_3PbI_3$ and special care is necessary for interpretation of earlier simulation results [5,6,109-113].

In Fig. 14(d), on the other hand, the experimental EQE spectrum shows the small reduction in the short wavelength region ($\lambda < 500$ nm), compared with the calculated EQE. This result suggests that the carrier recombination occurs slightly at the $TiO_2/CH_3NH_3PbI_3$ interface region. Moreover, the EQE calculated from our $CH_3NH_3PbI_3$ optical constants shows slightly smaller values particularly at $\lambda > 600$ nm, even though the EQE response agrees quite well at $\lambda = 500-600$ nm. We attributed the disagreement observed at $\lambda > 600$ nm to the enhanced carrier generation by the light scattering, which is not assumed in our optical simulation. In particular, as confirmed from Fig. 13, the carriers are generated uniformly at $\lambda > 600$ nm due to the low $\alpha$ in this regime and light absorption is more affected by the increase in the optical pass length. In Fig. 14(d), however, the difference of the EQE at $\lambda > 600$ nm is small and this contribution on the $J_{sc}$ (0.6 mA/cm$^2$) accounts for only 3.5% of the total $J_{sc}$. In high-efficiency tandem-type solar cells, however, it may be necessary to employ textured structures to improve the EQE response in the longer wavelength region. In reported $CH_3NH_3PbI_3$ solar cells, the EQE values often show rapid reduction at $\lambda > 500$ nm [4,12,15,49,54–57]. This trend can be interpreted by the weak optical confinement or shorter diffusion length of generated carriers in the solar cell.

## VI. DISCUSSION

All the optical transitions in $CH_3NH_3PbI_3$ are explained by the non-excitonic interband transitions within the $PbI_3^-$ cage. This phenomenon supports direct free-carrier generation at $E \geq E_g$. For the $E_0$ transition, the non-excitonic character has already been confirmed [5,66,67]. The $\alpha$ values determined from our analysis, however, show smaller values than those reported earlier [see Fig. 4(b)]. Figure 15 compares the $\alpha$ spectrum of $CH_3NH_3PbI_3$ in Fig. 3(b) to those of selected solar cell materials ($CuInSe_2$ [82], $CuGaSe_2$ [82], CdTe [82] and $c$–Si [114]). Although comparison of $\alpha$ spectra for various semiconductors has already been made [13,115,116], the previous $\alpha$ spectra are overestimated and we compare the $\alpha$ spectrum of $CH_3NH_3PbI_3$ obtained in this study to those of other semiconductors in Fig. 15. Rather surprisingly, the $\alpha$ values of $CH_3NH_3PbI_3$ are comparable to those of CIGS and CdTe absorbers, even though $J_{sc}$ of 20 mA/cm$^2$ is obtained with a thin layer thickness of 400 nm.

As confirmed from the results in Figs. 12(c) and 13, the high $J_{sc}$ can be attributed partly to effective back-side reflection realized by a $CH_3NH_3PbI_3$/transparent layer (*spiro*−OMeTAD)/metal (Ag or Au) structure. Thus, our results demonstrate the importance of the optical confinement particularly in the region of $E_0 < E < E_1$ (490 < $\lambda$ < 770 nm), where the $\alpha$ values are relatively low. In planar-type solar cells, the stronger optical confinement can be realized by a textured structure that enhances the optical pass length within the absorber layer while reducing the front light reflection. We find that, in the hybrid solar cells, the parasitic light absorption in the component layers is suppressed quite well. As known widely, CdS layers in CIGS solar cells [100,117] and a-Si:H p-type layers in a-Si:H/$c$–Si heterojunction solar cells [118,119] are "dead layers" that show strong parasitic absorption and the $J_{sc}$ reduces by 2 mA/cm$^2$ by these layers, whereas the light absorption in the $TiO_2$ front layer is quite weak with an optical loss of only 0.3 mA/cm$^2$ in Fig. 11.

As a result, the high performance of $CH_3NH_3PbI_3$ hybrid solar cells can be interpreted as being due to a combination of (a) efficient free carrier generation by semiconductor-type transitions within the $PbI_3^-$ component in the whole visible region ($E$ < 5 eV), (b) $\alpha$ values comparable to those of CIGS and CdTe, (c) high optical confinement and low parasitic absorption in the solar cells, (d) long-range carrier diffusion originating from low levels of carrier recombination in the interface and bulk regions, which has also been reported previously [11,21–23], (e) $E_g$ being close to the optimum value of 1.4 eV [120], and (f) the sharp absorption onset near $E_g$ (low $E_U$), which is expected to reduce the open-circuit voltage loss [13]. The absence of the gap state formation in $CH_3NH_3PbI_3$ [29−31] is another important factor in understanding of the superior solar cell characteristics.

## VII. CONCLUSION

The artifact-free dielectric function of $CH_3NH_3PbI_3$ is determined from ultra-smooth $CH_3NH_3PbI_3$ layers, fabricated by a laser evaporation technique, using SE. For the accurate optical characterization of $CH_3NH_3PbI_3$, the SE spectra of $CH_3NH_3PbI_3$ layers are obtained without exposing the samples to air and a self-consistent SE analysis is performed using the samples with different layer thicknesses. We find that the optical constants reported earlier for $CH_3NH_3PbI_3$ are influenced strongly by (i) the extensive surface roughness of $CH_3NH_3PbI_3$ samples and (ii) the formation of a hydrate phase near the surface region. The high $\alpha$ values deduced in previous $CH_3NH_3PbI_3$ studies are overestimated seriously due to a simple optical analysis performed for quite rough $CH_3NH_3PbI_3$ structures. Our SE analysis reveals that $\alpha$ is $3.8 \times 10^4$ cm$^{-1}$ at 2.0 eV and the $\alpha$ values near the $E_g$ are comparable to those of CIGS and CdTe semiconductors. The CP analysis of the $CH_3NH_3PbI_3$ dielectric function shows that the $E_g$ of $CH_3NH_3PbI_3$ is $1.61 \pm 0.01$ eV.

The optical transitions in $CH_3NH_3PbI_3$ are investigated further based on DFT calculations assuming the simple pseudo-cubic structure. We find that the optical anisotropy confirmed in the DFT-derived $CH_3NH_3PbI_3$ dielectric function arises from the strong interaction of $CH_3NH_3^+$ with the $PbI_3^-$ cage. In particular, the anisotropic optical properties vary significantly depending on the N atom position within the Pb–I network. When the effect of the center cation molecule is eliminated, the $CH_3NH_3PbI_3$ dielectric function deduced from DFT exhibits remarkable overall agreement with that extracted experimentally. To assign the optical transitions of $CH_3NH_3PbI_3$ in the visible/ultraviolet region, the dielectric response of each interband transition is calculated. As a result, the critical points observed at 2.53 eV and 3.24 eV in $CH_3NH_3PbI_3$ dielectric function are assigned to the direct optical transitions at the M and X points in the pseudo-cubic Brillouin zone, respectively.

The optical simulation of hybrid perovskite solar cells is also performed using the revised $CH_3NH_3PbI_3$ optical constants. Our optical simulation reproduces the EQE spectrum of a mesoporous $TiO_2$–$CH_3NH_3PbI_3$ solar cell fabricated by a standard solution process quite well, confirming the validity of the optical data obtained in this study. In the hybrid perovskite solar cell, the parasitic absorption induced by the solar-cell component layers is quite small. The partial IQE obtained from the optical simulation reveals that carriers are generated uniformly throughout the $CH_3NH_3PbI_3$ layer in the longer-wavelength region ($\lambda > 600$ nm) because of the low $\alpha$ values in this region. Accordingly, strong optical confinement and long carrier diffusion lengths are necessary to gain sufficient EQE response in this $\lambda$ region.

**TABLE I.** Tauc-Lorentz parameters extracted from the dielectric functions of the single layers shown in Fig. 10. The Tauc-Lorentz peaks shown in Fig. 10 correspond to the peaks summarized in this table.

| Material | Peak | $E_p$ (eV) | $A$ (eV) | $C$ (eV) | $E_g$ (eV) | $\varepsilon_1(\infty)$ |
|---|---|---|---|---|---|---|
| $CH_3NH_3PbI_3$ | 1[a] | 1.593 | 1.621 | 0.024 | 1.565 | 0 |
| | 2 | 1.607 | 47.677 | 0.137 | 1.593 | 1.486 |
| | 3 | 2.553 | 2.909 | 0.349 | 1.684 | 0 |
| | 4 | 3.046 | 39.202 | 2.391 | 1.563 | 0 |
| | 5 | 3.278 | 4.955 | 0.329 | 1.764 | 0 |
| | 6 | 3.581 | 1.369 | 0.354 | 2.246 | 0 |
| | 7 | 4.726 | 1.622 | 0.745 | 2.823 | 0 |
| | 8 | 5.648 | 4.137 | 0.666 | 3.839 | 0 |
| | 9 | 7.408 | 11.256 | 2.053 | 2.276 | 0 |
| $TiO_2$ | 1 | 3.904 | 94.341 | 0.693 | 3.215 | 0.537 |
| | 2 | 4.359 | 177.870 | 1.127 | 3.636 | 0 |
| | 3 | 5.514 | 28.662 | 1.028 | 4.894 | 0 |
| | 4 | 8.603 | 153.220 | 11.861 | 4.950 | 0 |
| *spiro*−OMeTAD | 1 | 3.102 | 55.872 | 0.283 | 2.855 | 2.256 |
| | 2 | 3.358 | 4.914 | 0.303 | 2.826 | 0 |
| | 3 | 4.035 | 3.033 | 0.555 | 2.850 | 0 |
| | 4 | 5.183 | 4.482 | 0.001 | 2.850 | 0 |
| Ag[b] | 1 | 2.560 | 1.711 | 4.950 | 0.0001 | 2.278 |
| | 2 | 3.506 | 0.045 | 0.417 | 0.267 | 0 |
| | 3 | 3.911 | 344.810 | 0.791 | 3.739 | 0 |
| $SnO_2$:F[c] | 1 | 7.000 | 44.550 | 12.000 | 3.300 | 2.575 |
| $PbI_2$ | 1 | 1.820 | 1.087 | 0.794 | 1.720 | 1.812 |
| | 2 | 2.462 | 162.630 | 0.124 | 2.374 | 0 |
| | 3 | 2.644 | 38.741 | 0.415 | 2.335 | 0 |
| | 4 | 2.902 | 55.440 | 0.453 | 2.316 | 0 |
| | 5 | 3.208 | 61.819 | 0.200 | 3.023 | 0 |
| | 6 | 3.829 | 75.977 | 1.428 | 2.211 | 0 |
| | 7 | 4.364 | 24.014 | 0.441 | 3.376 | 0 |
| | 8 | 5.313 | 3.436 | 1.381 | 1.800 | 0 |

**TABLE I.** (*Continued.*)

| Material | Peak | $E_p$ (eV) | $A$ (eV) | $C$ (eV) | $E_g$ (eV) | $\varepsilon_1(\infty)$ |
|---|---|---|---|---|---|---|
| $CH_3NH_3I$ | 1 | 3.743 | 2.018 | 1.556 | 2.200 | 1.405 |
| | 2 | 4.601 | 1.994 | 0.547 | 2.200 | 0 |
| | 3 | 5.302 | 0.967 | 0.444 | 2.200 | 0 |
| | 4 | 6.083 | 0.608 | 0.671 | 2.200 | 0 |

[a] Peak 1 has been included to improve the fitting in the $\alpha$ spectrum in the energy region below $E_g$.

[b] Drude parameters for $\varepsilon = -A_D/(E^2 - i\Gamma_D E)$ are $A_D = 81.448$ eV and $\Gamma_D = 0.040$ eV.

[c] Modeled Tauc-Lorentz parameters are shown. The Drude parameters for $\varepsilon = -A_D/(E^2 - i\Gamma_D E)$ are $A_D = 0.919$ eV and $\Gamma_D = 0.129$ eV. These parameters correspond to $N_{opt} = 2 \times 10^{20}$ cm$^{-3}$ and $\mu_{opt} = 30$ cm$^2$/(Vs).

**Figure Captions**

FIG. 1. (a) Schematic of the laser evaporation process, (b) SEM image of the $CH_3NH_3PbI_3$/ZnO/$c$−Si structure, (c) XRD spectra of the $CH_3NH_3PbI_3$/ZnO/$c$−Si and $PbI_2$/$c$−Si structures and (d) AFM images of the samples analyzed by SE. In (d), the thickness of each layer (value in parenthesis) and the root-mean-square roughness ($d_{rms}$) obtained from each AFM image are also indicated.

FIG. 2. ($\psi$, $\Delta$) ellipsometry spectra obtained from the $CH_3NH_3PbI_3$ (85 nm)/ZnO (50 nm)/$SiO_2$ (2 nm)/$c$−Si structure. The solid lines show the fitting result calculated using the $CH_3NH_3PbI_3$ dielectric function extracted from the thinner layer (45 nm).

FIG. 3. (a) Dielectric functions and (b) $\alpha$ spectra of the $CH_3NH_3PbI_3$, $PbI_2$ and $CH_3NH_3I$ layers, (c) $E_g$ analysis for $CH_3NH_3PbI_3$ using the $(\alpha E)^2$–$E$ plot, and (d) CP analysis of $CH_3NH_3PbI_3$. In (a) and (b), the dotted lines for $CH_3NH_3PbI_3$ are obtained from the measurement in air. In (c), the $E_g$ analysis results for two different $(\alpha E)^2$ regions are shown. The open circles represent the experimental data and the solid lines represent the linear fitting results. In (d), the open circles denote the experimental data and the solid line represents the theoretical fitting. The CP energies determined from the analysis are indicated by the arrows.

FIG. 4. (a) $\varepsilon_2$ spectra [5–10] and (b) $\alpha$ spectra [5–13] of $CH_3NH_3PbI_3$ reported earlier in other studies, (c) change of the $CH_3NH_3PbI_3$ $\varepsilon_2$ spectrum in air at 40% relative humidity, and (d) variation of the pseudo-$\varepsilon_2$ spectrum ($\langle\varepsilon_2\rangle$) and the pseudo-$\alpha$ spectrum ($\langle\alpha\rangle$) with the roughness layer thickness, calculated from the $CH_3NH_3PbI_3$ dielectric function determined in this study using an optical model of ambient/surface roughness/ $CH_3NH_3PbI_3$ substrate. In (a) and (b), the $\varepsilon_2$ and $\alpha$ spectra obtained from our analysis [Figs. 3(a) and (b)] are also shown. In (b), the $\alpha$ spectra denoted by the dotted lines represent those that show persistent $\alpha$ values in the energy region below $E_g$. In (c), the $\varepsilon_2$ spectrum denoted as "in $N_2$" corresponds to the one shown in Fig. 3(a). In (d), when the roughness thickness is zero, $\langle\varepsilon_2\rangle$ and $\langle\alpha\rangle$ are equivalent to $\varepsilon_2$ and $\alpha$, respectively. In (b) and (d), the $E_g$ position ($E_g$ = 1.61 eV), determined by the CP analysis in Fig. 3(d), is indicated.

FIG. 5. (a) XRD, (b) SEM, (c) dielectric function, and (d) $\alpha$ spectra of the $CH_3NH_3PbI_3$ layers annealed at 80 °C and 100 °C for 45 min in vacuum without the $CH_3NH_3I$ flux.

The results for the as-deposited $CH_3NH_3PbI_3$ layer shown in Figs. 1 and 3 are also indicated.

FIG. 6. (a) Pseudo-cubic $CH_3NH_3PbI_3$ crystal structure obtained from the DFT calculation, (b) $\varepsilon_2$ spectra of $CH_3NH_3PbI_3$, obtained from the experiment (open circles) and the theoretical calculation (solid lines), (c) band structure and DOS of the pseudo-cubic $CH_3NH_3PbI_3$ crystal, and (d) high symmetry points in the Brillouin zone for the pseudo-cubic structure. In (b), the $\varepsilon_2$ spectra in two different energy regions are shown. These $\varepsilon_2$ spectra are calculated assuming different polarization states parallel to the *a*, *b*, and *c* axes indicated in (a) and the CP energies that correspond to the optical transitions in (c) are indicated by arrows. In (c), the optical transitions in $CH_3NH_3PbI_3$, assigned by the polarization-dependent DFT analysis, are indicated by arrows. The insets of (c) show the charge density profiles for the energy regions of $E = -0.2$–$0.0$ eV (VBM), $E = 1.4$–$1.6$ eV (CBM) and $E = 2.0$–$2.2$ eV. For the density profile of $E = 2.0$–$2.2$ eV, the charge localization on the N atom is indicated by the dotted circle. Partial DOS distributions of the Pb $6s$, Pb $6p$ and I $5p$ states are also shown.

FIG. 7. Contribution of various interband transitions to the $\varepsilon_2$ spectrum of $CH_3NH_3PbI_3$, determined from DFT calculations: $\varepsilon_2$ spectra for different polarization states along the (a) *a* axis, (b) *b* axis, (c) *c* axis, and (d) $\varepsilon_2$ spectra calculated from $(\varepsilon_a + \varepsilon_b + \varepsilon_c)/3$. $V_1C_2$ indicates the $\varepsilon_2$ contribution induced by the optical transition from the first valence band ($V_1$) to the second conduction band ($C_2$) in Fig. 6(c), for example. The $\varepsilon_2$ spectra denoted as "Total" in (a)–(c) correspond to the $\varepsilon_2$ spectra shown in Fig. 6(b). Only the transitions between $V_j$ and $C_j$ ($j \leq 4$) with a peak amplitude of $\varepsilon_2 > 0.5$ are shown for clarity.

FIG. 8. Dielectric functions of $NH_4PbI_3$ and $CH_3NH_3PbI_3$ obtained from DFT calculations: (a) $NH_4PbI_3$ with the N atom located at the center position of the C–N bond in the structure of Fig. 6(a), (b) $NH_4PbI_3$ after the structural relaxation of $NH_4^+$, and (c) $CH_3NH_3PbI_3$ with the Pb–I–Pb bond angle of 180° in the structure of Fig. 6(a). In the DFT calculations, the atomic configuration of the $PbI_3^-$ cage is assumed to be identical and is represented by that of the optimized $CH_3NH_3PbI_3$ structure in Fig. 6(a), except for (c). The insets show the crystal structures used for the DFT calculations.

FIG. 9. (a) Comparison of the $CH_3NH_3PbI_3$ $\varepsilon_2$ spectrum calculated in this study with those reported in other studies [32–35] and (b) DOS distributions of the pseudo-cubic

and tetragonal structures. In (a), the DFT results are shown by the lines, whereas the experimental dielectric function obtained from our analysis is indicated by the open circles. The $\varepsilon_2$ spectrum in this study corresponds to $\varepsilon_b$ in Fig. 6(b). In the previous studies, the DFT calculations were performed assuming the cubic structure by PBE [32] and tetragonal structure by HSE06 [35]. The LDA has also been applied assuming the tetragonal [33] and orthorhombic [34] structures. For these studies, the $\varepsilon_2$ spectra for the $a$ axis component are shown. The experimental CP energies, determined from the analysis of Fig. 3(d), are also indicated.

FIG. 10. Dielectric functions of the solar-cell component layers ($CH_3NH_3PbI_3$, $TiO_2$, *spiro*–OMeTAD, and Ag), together with the $PbI_2$ and $CH_3NH_3I$ layers, extracted from the SE analyses. The open circles show the experimental dielectric functions and the solid lines indicate the fitting results obtained using the Tauc-Lorentz model. For the dielectric function modeling of Ag, the Drude model is also used. The model parameters are summarized in Table I.

FIG. 11. Calculated $A$ spectra of the component layers and EQE spectrum for the hybrid perovskite solar cell. The reflectance spectrum ($R$) of the solar cell is also shown. The optical model of the solar cell is indicated in the inset. The $A$ spectrum of $CH_3NH_3PbI_3$ corresponds to the EQE spectrum when a carrier collection efficiency of 100% is assumed.

FIG. 12. (a) IQE spectrum of the $CH_3NH_3PbI_3$ layer and IA spectra of the component layers in the glass/$SnO_2$:F (600 nm)/$TiO_2$ (200 nm)/$CH_3NH_3PbI_3$ (400 nm)/*spiro*–OMeTAD (500 nm)/Ag structure, (b) normalized partial IQE spectra for the $CH_3NH_3PbI_3$ and partial IA spectra for the $TiO_2$, and (c) integrated $J_{sc}$ for perovskite solar cells with discrete $CH_3NH_3PbI_3$ thicknesses from 100 to 1000 nm. For the $CH_3NH_3PbI_3$ layer in (b), integrated $J_{sc}$ values relative to the depth from the $TiO_2$ interface and $\lambda$ are shown. The dotted lines in this figure show the $E_0$ ($E_g$) positions. In (c), the solid circles denote the end points of the $CH_3NH_3PbI_3$ thickness and the dotted line shows the $J_{sc}$ values obtained by varying the $CH_3NH_3PbI_3$ layer thickness in the solar cell.

FIG. 13. Normalized partial IQE calculated for different depths from the $TiO_2$/$CH_3NH_3PbI_3$ interface and wavelengths. The solar cell structure is identical to that of Fig. 12, but the $CH_3NH_3PbI_3$ layer thickness is varied (200, 400 and 600 nm). The simulation result for the 400-nm-thick layer corresponds to the result shown in Fig. 12(b). The integrated $J_{sc}$ values relative to the depth from the $TiO_2$ interface are shown in Fig. 12(c).

FIG. 14. (a) Optical model constructed for a $CH_3NH_3PbI_3$ solar cell consisting of $MgF_2$/glass/$SnO_2$:F/compact $TiO_2$ (30 nm)/mesoporous $TiO_2$–$CH_3NH_3PbI_3$ (150 nm)/$CH_3NH_3PbI_3$ (240 nm)/*spiro*–OMeTAD/$MoO_x$/ZnO:Al/(Ni–Al grid)/$MgF_2$ reported in [101], (b) the optical constants of the $MgF_2$, $SnO_2$:F, $MoO_x$, and ZnO:Al used in the optical simulation, (c) the $T$ and $R$ spectra obtained experimentally from the semi-transparent solar cell ($T_{ex}$, $R_{ex}$: open circles) [101], together with the simulated $T$ spectra and (d) the EQE spectrum obtained experimentally from the semi-transparent solar cell (open circles) [101], together with the simulated EQE spectra. In (b), the ($n$, $k$) data of the $MgF_2$, $SnO_2$:F, and ZnO:Al are taken from Ref. [100], Ref. [102] and Ref. [100], respectively. In (c) and (d), the solid lines represent the $T$ and EQE spectra calculated assuming different $CH_3NH_3PbI_3$ optical constants obtained in this study (red line), Ref. [6] (green line) and Ref. [8] (blue line). The $J_{sc}$ values calculated from the EQE spectra are also indicated.

FIG. 15. $\alpha$ spectra of various solar cell materials. The $\alpha$ spectrum of $CH_3NH_3PbI_3$ corresponds to the one shown in Fig. 3(b). In this figure, the $\alpha$ spectra of $CuInSe_2$ [82], CdTe [82], $CuGaSe_2$ [82] and *c*–Si [114] are also shown.

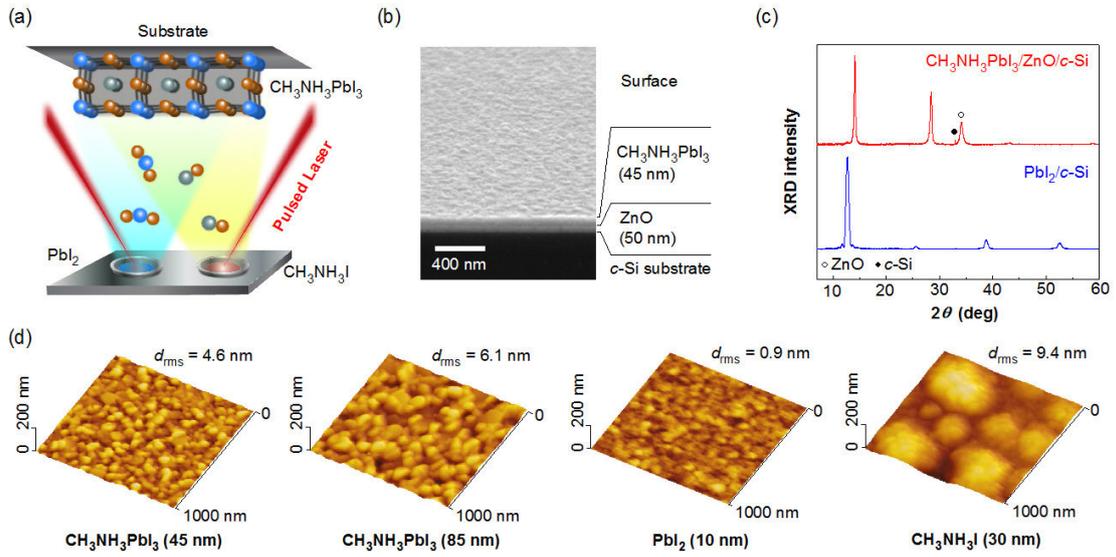

**Figure 1**

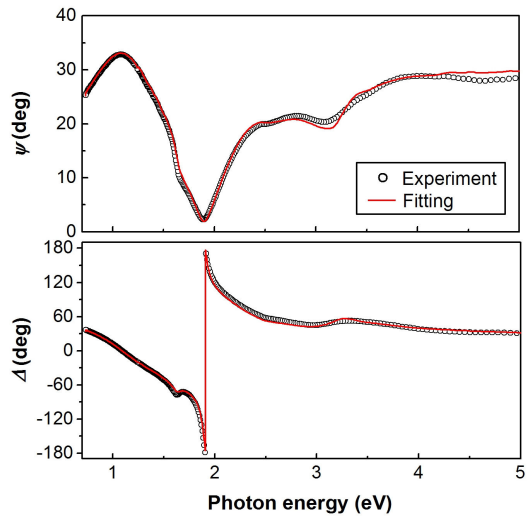

**Figure 2**
46

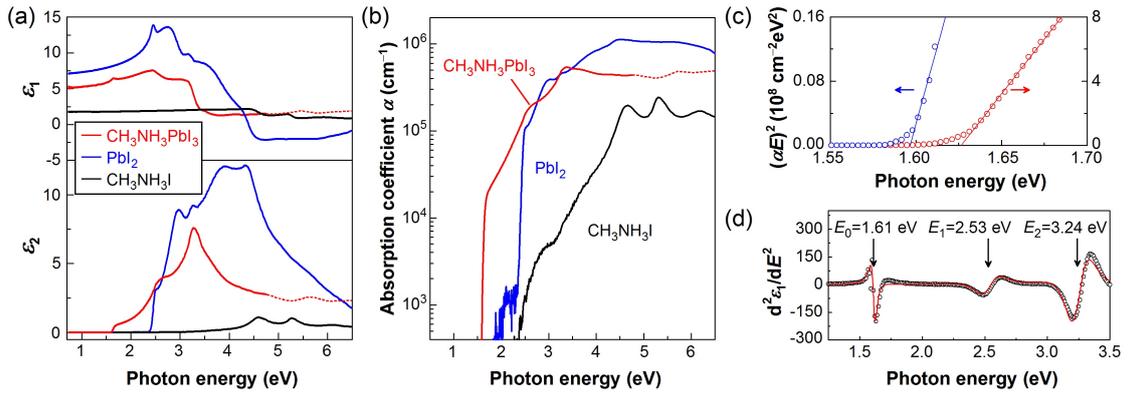

Figure 3

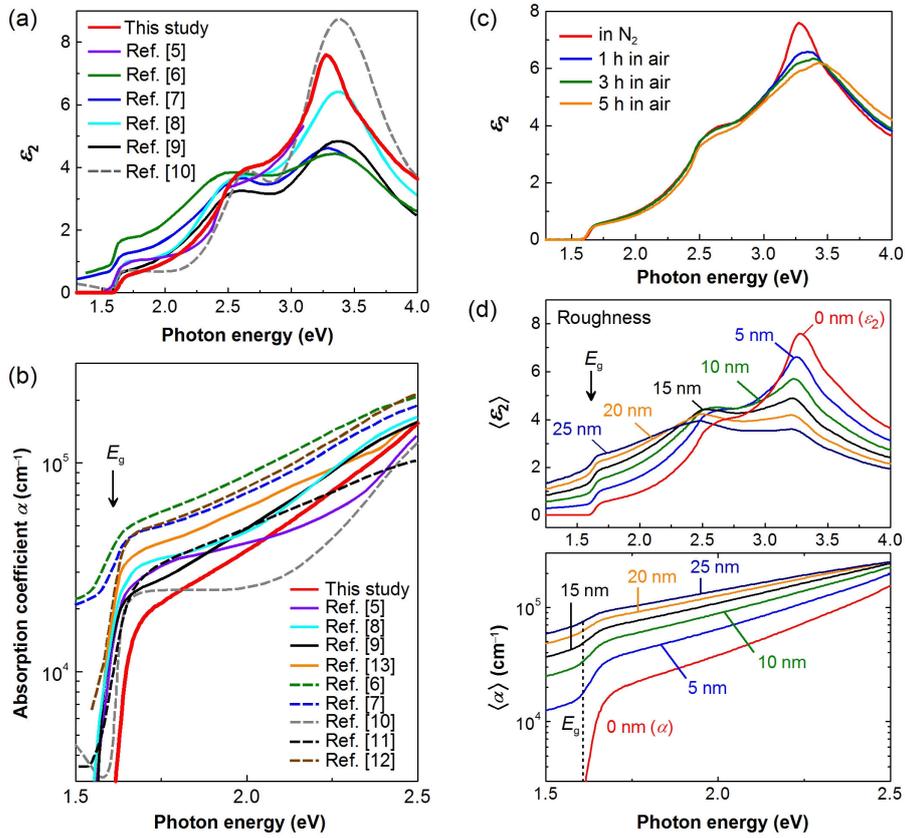

Figure 4

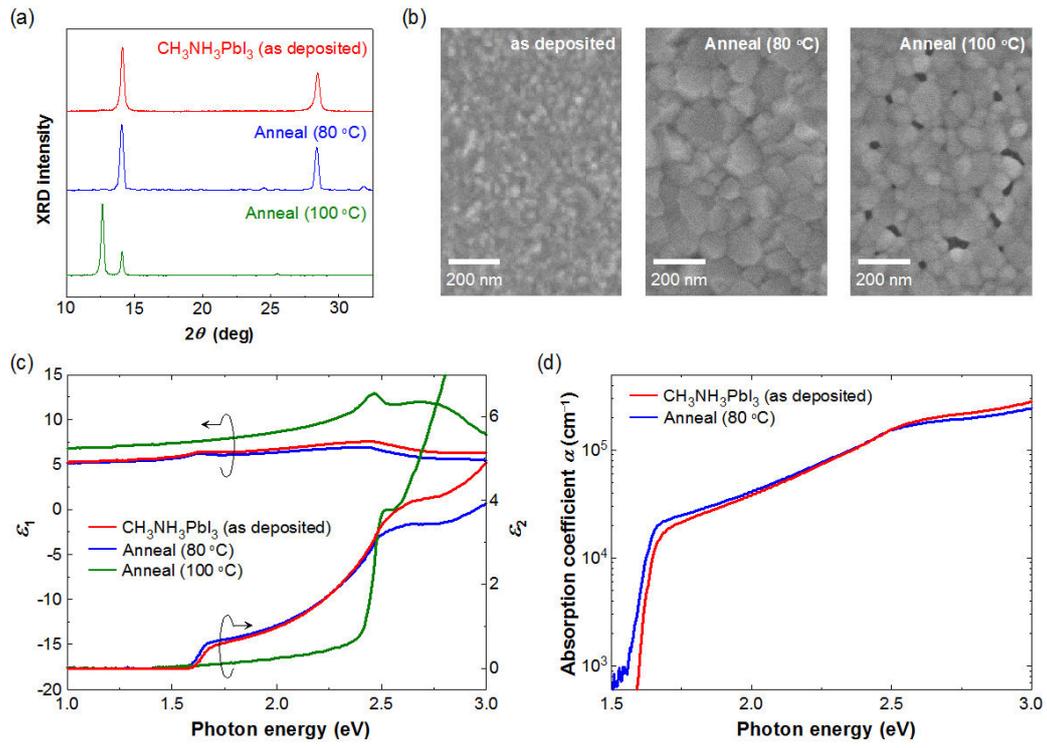

Figure 5

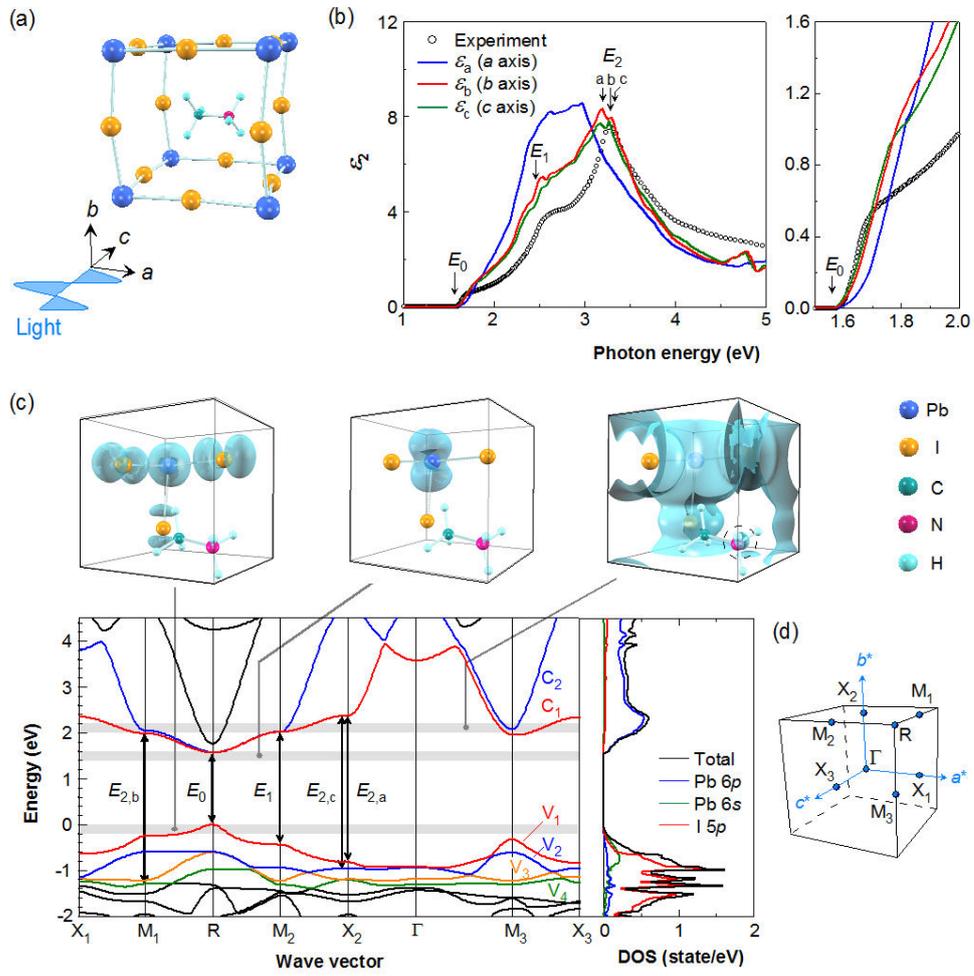

**Figure 6**

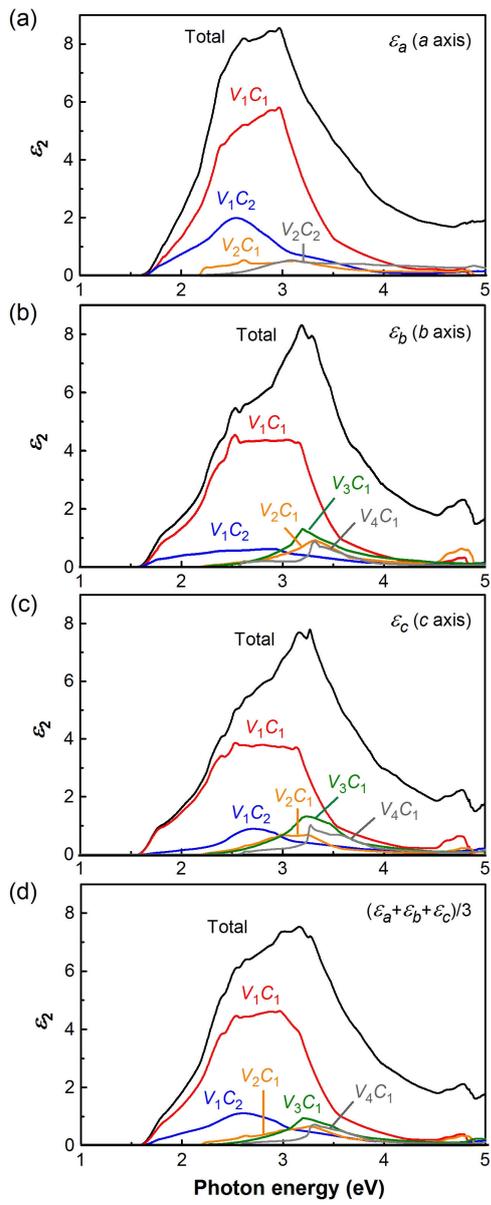

**Figure 7**

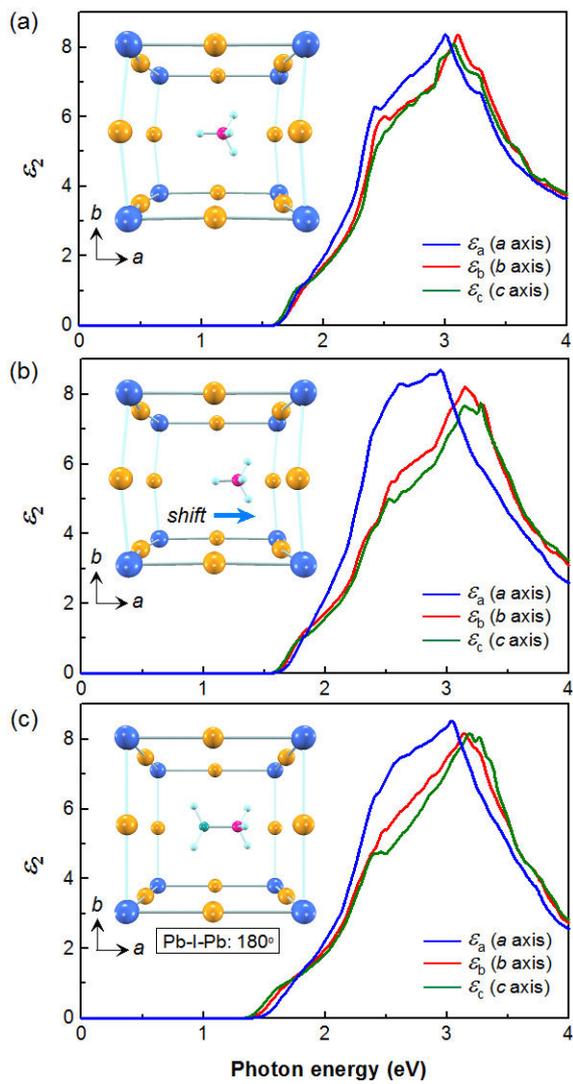

**Figure 8**

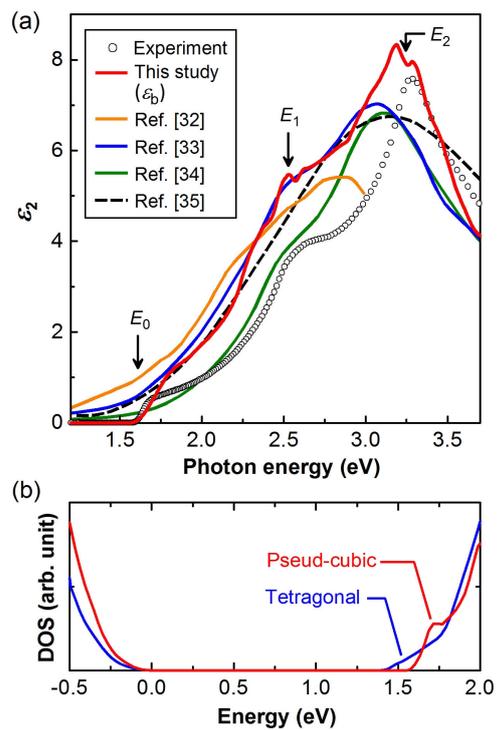

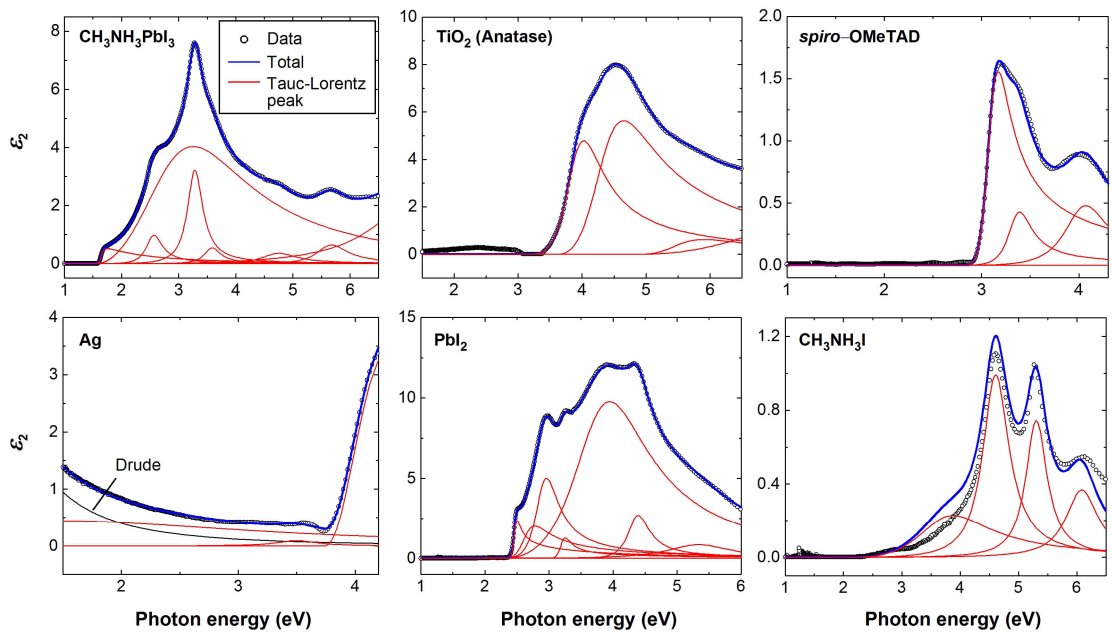

Figure 10

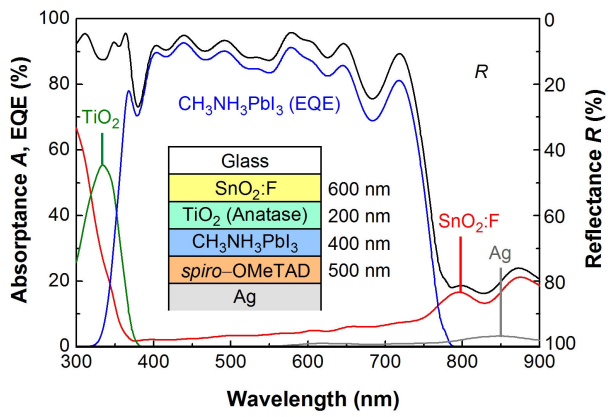

Figure 11

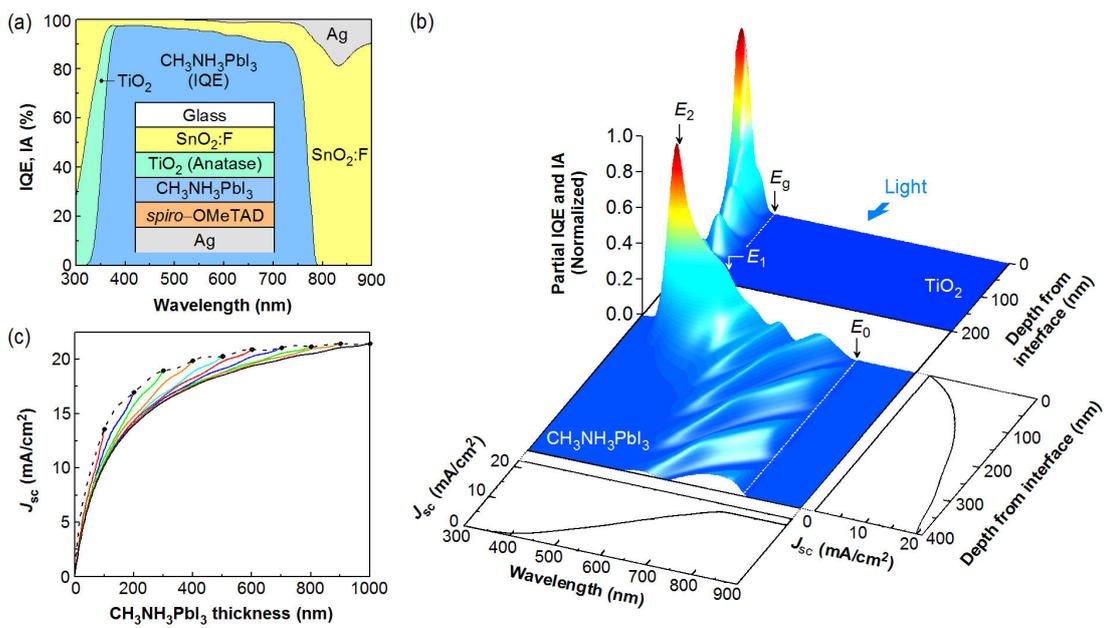

Figure 12

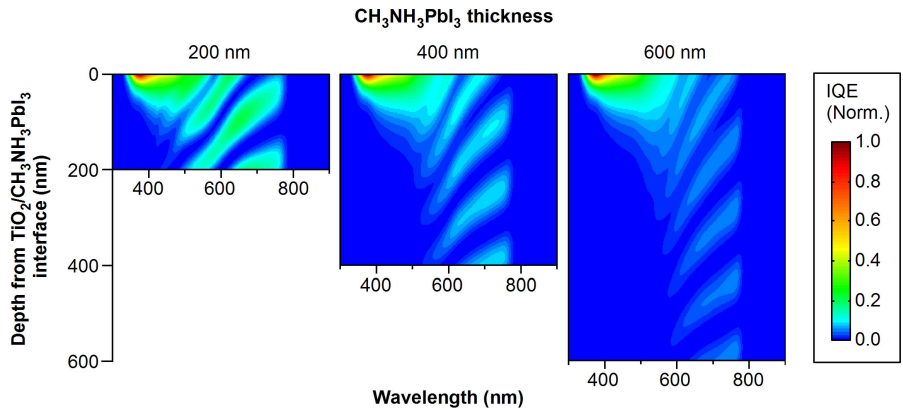

Figure 13

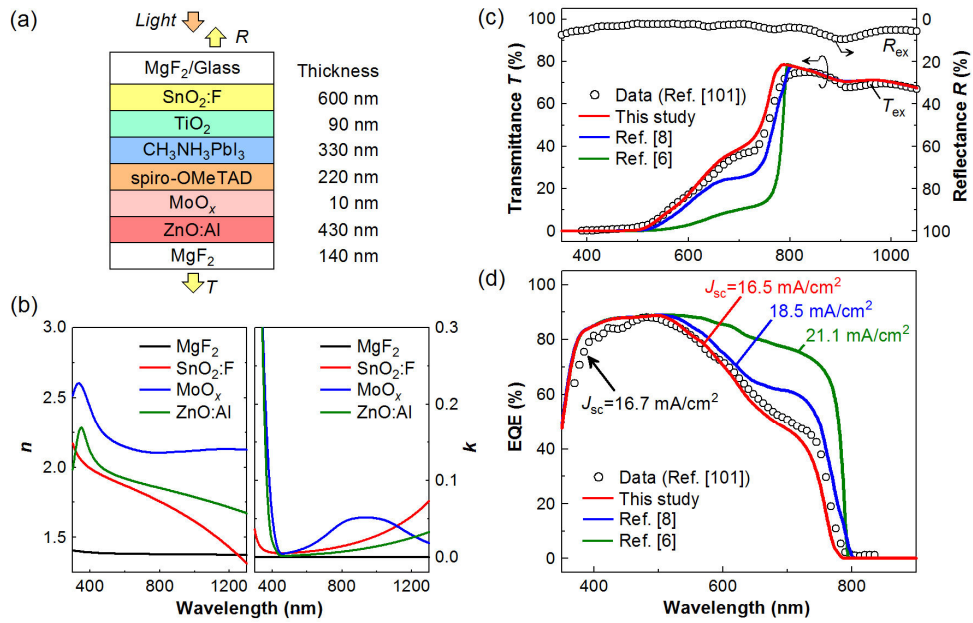

Figure 14

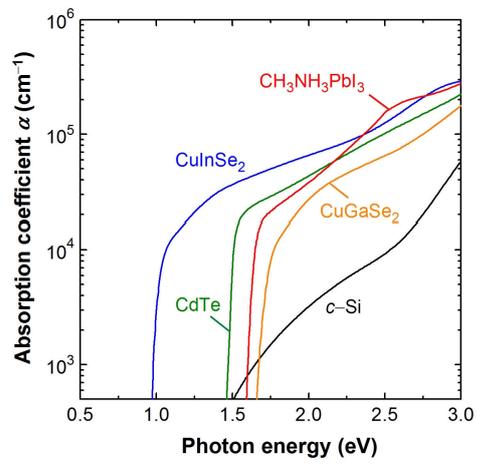

**Figure 15**